# Two-dimensional electronic spectroscopy of organic semiconductor nanostructures

*Daniel Timmer[1], and Christoph Lienau[1,2,3,\*]*

[1]Institut für Physik, Carl von Ossietzky Universität, Carl-von-Ossietzky Str. 9-11, 26129 Oldenburg, Germany

[2]Center for Nanoscale Dynamics (CeNaD), Carl von Ossietzky Universität, Carl-von-Ossietzky Str. 9-11, 26129 Oldenburg, Germany

[3]Research Centre for Neurosensory Sciences, Carl von Ossietzky Universität, Carl-von-Ossietzky Str. 9-11, 26129 Oldenburg, Germany

[\*] Correspondence to: christoph.lienau@uni-oldenburg.de

## Abstract

This chapter discusses recent experimental work exploring the optical properties and quantum dynamics of organic semiconductor nanostructures based on squaraine dyes. Squaraines are prototypical quadrupolar charge-transfer chromophores of interest for solution-processed photovoltaics and as aggregates with large circular dichroism. Here, we demonstrate and exploit their unique properties as quantum emitters for implementing hybrid nanostructures featuring strong exciton-plasmon couplings. We show that the unusual electronic properties of squaraines result in a substantial reduction of vibronic coupling to the ubiquitous high-frequency C-C-bond-stretching modes of organic materials and in the formation of spectrally narrow J-aggregated exciton resonances in squaraine thin films. This is exploited to create metallic nanostructures covered with squaraine thin films and to perform the first time-domain study of coherent exciton-plasmon couplings using two-dimensional electronic spectroscopy with 10-fs time resolution. The experiments present unexpected evidence for long-range coherent exciton transport driven by plasmonic fields. This opens up new opportunities for manipulating the coherent transport of matter excitations by coupling to vacuum fields.

1. Introduction

Squaraines[1] are prototypical quadrupolar charge-transfer molecules comprising two nitrogen-containing donors (D) that are connected via a squaric acid acceptor (A) by $\pi$-conjugated bridges (Fig. 1). Such quadrupolar D-$\pi$-A-$\pi$-D molecules are of substantial current interest because of their large transition dipole moments, resulting from the delocalized $\pi$ electrons, and their interesting nonlinear optical properties.[2, 3] Squaraines exhibit a narrow long-wavelength absorption band, with a high absorption coefficient on the order of 300000 L mol$^{-1}$ cm$^{-1}$, and comparatively high fluorescence quantum yields, which makes them good near IR absorbers and emitters.[4] Squaraines are also promising as light absorbers and donor materials for efficient organic photovoltaic devices,[5-7] and can improve the efficiency of polymer solar cells as an additive.[8] Furthermore, their comparatively large two-photon absorption cross sections[9] make them interesting for fluorescence bioimaging and photopolymerization applications.[10, 11] Synthetic chemical approaches can readily be used to adapt the electron donating or accepting properties of the molecules and the molecular arrangement in squaraine aggregates [5, 12-14] This has recently led to the observation of strong circular dichroism in helically-ordered thin films of aggregated squaraine monomers.[15]

Much less known are the effects of the electronic structure of such quadrupolar charge-transfer dyes on couplings to molecular vibrations.[16, 17] Phenomenological essential state models

(ESM) for such quadrupolar dyes[2, 17] indicate that the formation of electronic states with strong charge transfer character and the interplay between charge delocalization across the molecular backbone and coupling to local vibrations may profoundly affect the quantum dynamics. This is expected both in the isolated molecule and in aggregated thin films, but such effects have not yet been studied experimentally.[2, 16-18] It is the aim of the this chapter to review and discuss recent experimental work that explores vibronic couplings in squaraine molecules[19] and couplings of excitons in aggregated squaraine thin films to plasmonic modes[20, 21] using advanced ultrafast pump-probe and two-dimensional electronic spectroscopy (2DES).[22, 23] The obtained results are exploited to design hybrid nanostructures comprising gold nanoslit arrays covered with J-aggregate squaraine thin films and, using 2DES with 10-fs time resolution, to perform the first time-domain study of coherent exciton-plasmon couplings in such systems.

The chapter is structured as follows. Section 2 briefly describes the high-repetition rate 2DES setup developed to perform these experiments. Section 3 gives a brief introduction into 2DES. Section 4 introduces the phenomenological ESM, adapts it to squaraine molecules, and compares the results to quantum-chemical calculations. It then discusses vibronic couplings in squaraine molecules. Section 5 introduces the optical properties of J-aggregated thin films of squaraines and analyzes the effects of couplings of J-aggregated squaraine excitons to plasmonic modes of a gold film. Section 6 summarizes the results of a 2DES study of strong exciton-plasmon coupling dynamics in squaraine/gold nanoslit hybrids using 2DES with 10-fs time resolution. The chapter ends with some conclusions and an outlook.

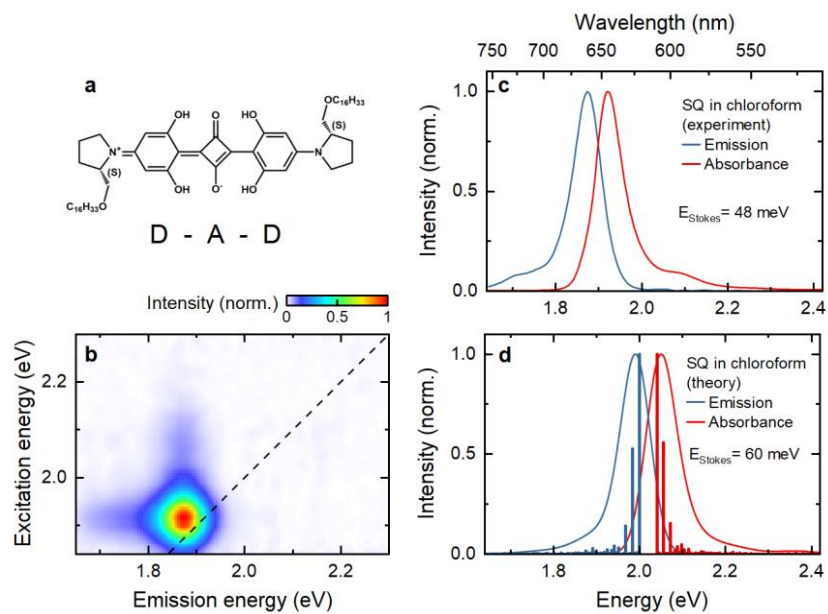

**Figure 1**: Linear spectroscopy of the donor-acceptor-donor (D-A-D) squaraine molecule ProSQ-C16 (SQ) in chloroform solution at room temperature. **a**) Chemical structure of the molecule. **b**) Photoluminescence (PL) excitation-emission map of SQ in chloroform, indicating a Stokes shift of 48 meV. **c**) Experimental absorption and emission spectra. **d**) Quantum-chemical simulations of the vibrationally resolved absorption and emission spectra for SQ in chloroform. Individual transitions and their oscillator strengths are shown as sticks. These figures are reprinted with permission from Ref. [19], Copyright (c) 2016 APS Journals.

## 2. Experimental setup for high-repetition rate 2DES with 10-fs time resolution

The main aim of the experimental work is to report on time-domain pump-probe and 2DES studies of vibronic couplings in squaraine molecules and of coherent exciton/plasmon couplings in squaraine/metal hybrid nanostructures[19]. In organic molecules, vibronic couplings to high-frequency carbon-carbon backbone modes with frequencies in the 1500 cm$^{-1}$ range are of functional relevance.[24-26] Their vibrational period of around 22 fs puts an upper limit on the time resolution of the experimental setup that is needed to temporally resolve the effect of the vibrational motion on the optical and electronic properties of the molecules. Also the dephasing times of plasmonic modes in metallic nanostructures typically lie in the range of few to few tens of femtoseconds.[27-29] Reaching the regime of strong exciton/plasmon (X-SPP) coupling therefore requires coupling strengths that exceed the inverse dephasing time.[30-32] This again calls for ultrafast studies with a time resolution in the range of 10 fs or even below to temporally resolve the oscillatory energy transfer processes that are underlying strong coupling. Even though pump-probe and 2DES are highly developed methods,[33] this is still at the limit of what is currently achieved experimentally. We therefore developed a setup that combines a high time resolution of better than 10 fs with short data acquisition times afforded by the use of a high-repetition rate laser system and a sensitive detection scheme. The heart of this setup is an inherently phase-stable common-path interferometer, the Translating Wedge-based Identical pulse eNcoding System[34] (TWINS), that has been demonstrated to be a simple and comparatively cost-effective solution for the generation of phase-locked excitation pulse pairs for 2DES.[34, 35] The working principle is illustrated in Fig. 2.

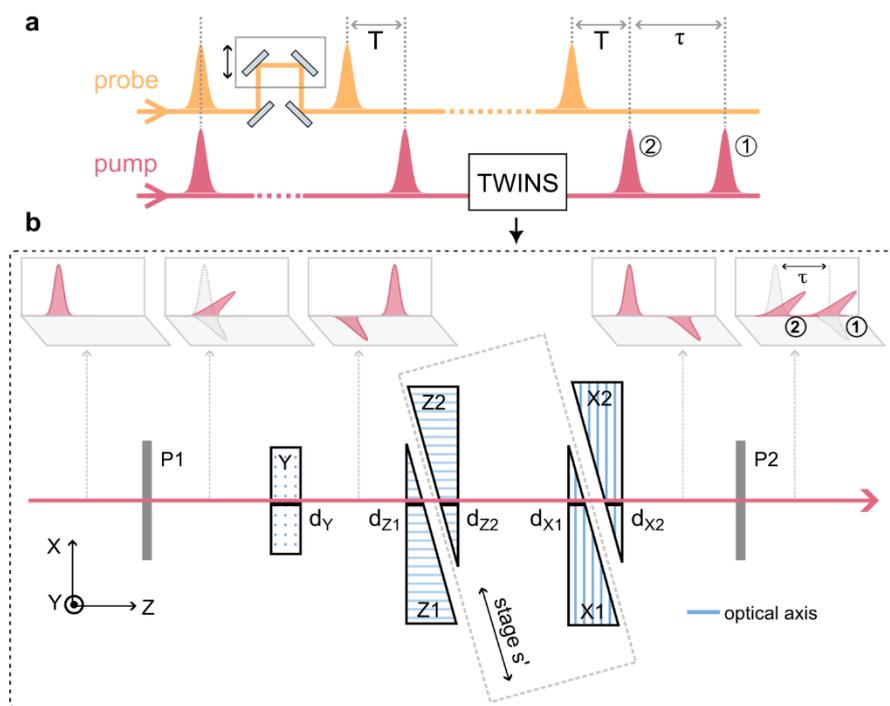

**Figure 2: a)** Pulse sequence used in the 2DES experiments. Pump and probe pulses are delayed by the waiting time $T$. A TWINS interferometer then creates a phase-stable pump pulse pair with delay $\tau$. **b)** Scheme of the TWINS for phase-cycling. A single input pump pulse, polarized at 0° relative to the vertical passes the first polarizer (P1) set to 45°. A birefringent $\alpha$-BBO crystal with optical axis along Y and two pairs of $\alpha$-BBO wedges with optical axes oriented along Z and X, respectively, split the input pump pulse and create a pump-pulse pair with tunable delay $\tau$. Translation of the $Z_2$ and $X_1$ wedges introduces a time delay of the first pulse (1). Both pulses (1) and (2) are set to collinear polarization using a second polarizer P2.

In the TWINS, the input pulse is split into a pair of s- and p-polarized beam components which propagate through two pairs of birefringent wedges with different orientations of their optical axis. This propagation results in a phase delay of $\Delta\phi = -\left(\frac{\omega}{c_0}\right)\Delta n\, d\, \sin\alpha$ between the two beams that is proportional to the birefringence of the wedge material, $\Delta n = n_o - n_e$, and the position $d$ of the motor stage that controls the second and third wedge in Fig. 2b. For BBO wedges with an opening angle of $\alpha = 7°$, the resulting group delay $\tau_g = \partial\Delta\phi(\omega)/\partial\omega$ is ~ 50 fs per mm of stage motion.[36, 37] After careful calibration of the motor stage,[37] the phase can be controlled with a precision of better than $\lambda/250$ and delay scans of up to ~ 1 ps can be performed at this level of precision within ~ 10 seconds.[36, 37]

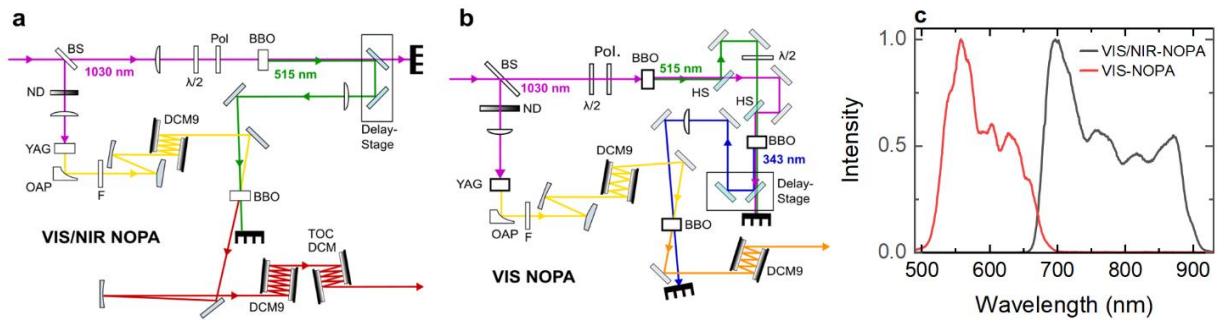

**Figure 3:** High-repetition rate non-collinear optical parametric amplifiers (NOPAs) **a)** Design of a VIS/NIR-NOPA, pumped at 515 nm and covering the spectral range from 650 to 900 nm. BS: beam splitter, $\lambda/2$: half-wave plate, Pol: polarizer, ND: neutral density filter, OAP: off-axis parabolic mirror, F: filter. **b)** Design of the VIS-NOPA, pumped at 343 nm and covering the spectral range from 500 to 800 nm. HS: harmonic separator. **c)** Typical output spectra obtained with the two NOPAs.

In the experiments described in this chapter, we use few-cycle pump pulses derived from a home-built high repetition rate noncollinear parametric amplifier (NOPA) system.[20, 38-41] The NOPAs are either pumped by an Yb-based fiber amplifier system operating at 175 kHz (Tangerine V2, Amplitude Systèmes), creating 260-fs pulses at 1030 nm or by a fraction of an 80-W-Yb:KGW laser (Light Conversion, Carbide),[36] delivering 190-fs pulses at 1030 nm at a repetition rate of 200 kHz. The design of the NOPA stages is based on that reported in Ref. [41] and briefly illustrated in Fig. 3.[20, 21, 23] When using the second harmonic of the pump laser for pumping, we generate VIS/NIR-NOPA pulses with a spectrum extending from ~650 to ~900 nm and a pulse energy of ~ 1 µJ (Fig. 3c). Using chirped mirrors (DCM9, Laser Quantum and custom-built, Laseroptik) for dispersion compensation, we reach pulse durations of ~ 10 fs (full width at half maximum of the intensity profile) after the TWINS interferometer (Fig. 4). When using the third harmonic of the pump laser (Fig. 2b), the NOPA can be designed to deliver ~ 1 µJ pulses with a spectrum covering the range from 500 to 700 nm (Fig. 3c). With chirp-mirror compression, their pulse duration can be reduced to less than 10 fs. When pumping the NOPAs with the fiber amplifier system, we reach shot-to-shot energy fluctuations of the pulses of ~ 1% rms. Reaching such noise levels requires careful adjustment of both the pump laser and the NOPA. Using the Yb:KGW laser as a pump, the shot-to-shot energy fluctuations are reduced to <0.4% rms. With such high pulse energy stability, our high-repetition rate detection scheme reaches a signal-to-noise ratio that exceeds the shot-noise limit by less than 50%. This is of crucial importance for reducing the acquisition time of the 2DES measurements.

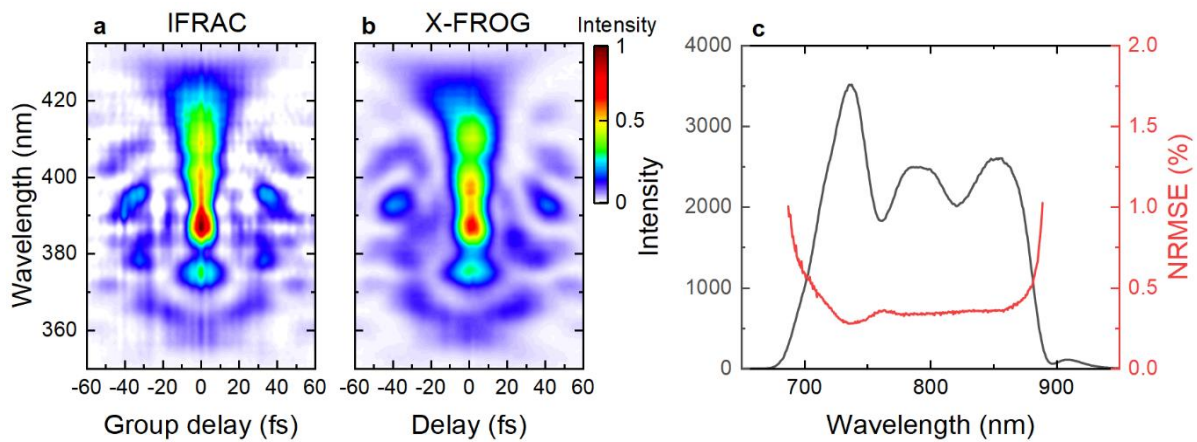

**Figure 4:** Laser pulse characterization. **a)** IFRAC measurement of the pump pulse using the TWINS interferometer. **b)** Cross-correlation SHG-FROG between pump and probe at the sample position. Both traces yield a retrieved pulse duration of 10 fs. **c)** Laser spectrum and single-shot RMS fluctuations.

The NOPA pulses are fed into the high-repetition rate 2DES spectrometer depicted schematically in Fig. 5. After the input beam is split into pump and probe pulses using a broadband beam splitter, the pump pulses are periodically switched on and off using a home-built chopper wheel at a rate of ¼ of the laser repetition rate, i.e., in pairs of two laser pulses. A phase-locked pair of collinearly propagating pump pulses with a delay that is denoted as the "coherence time" $\tau$ is generated in the TWINS interferometer. After chirped-mirror compression, the pump-pulses and the probe pulses are focused onto the sample under a small angle using an off-axis parabolic mirror. The time delay between the arrival of the second pump pulse and the probe pulse in the sample plane is denoted as the "waiting time" $T$ (Fig. 2a). If the waiting time is positive ($T > 0$), the probe arrives after the second pump pulse. Experimentally, the waiting time can be tuned using a retroreflector that is mounted on a motorized translation stage. A vibrating piezoelectric mirror (VM) in the probe beam is used to periodically modulate the pump-probe delay in order to suppress scattering contributions.[35] The light that is emitted in the probe direction is collected either in transmission or reflection geometry and spectrally dispersed in a monochromator. For detection, we employ a high-repetition rate line camera (e2v Aviivva EM4) with 1024 pixels and a maximum acquisition rate of 126 kHz. Spectra are therefore recorded at half the laser repetition rate. Since the pump pulses are chopped, every sequence of two spectra measures the probe transmission (or reflection) with and without prior pumping, $S_{on}(E_{det}, \tau, T)$ and $S_{off}(E_{det})$, respectively. From these, differential transmission (or reflection) spectra can be calculated as

$$\frac{\Delta T}{T}(E_{det}, T, \tau) = \frac{S_{on}(E_{det}, T, \tau) - S_{off}(E_{det})}{S_{off}(E_{det})} \qquad (1)$$

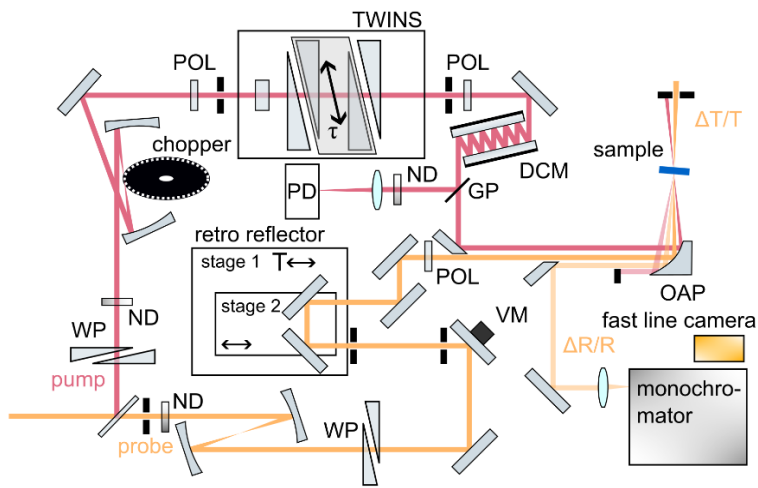

**Figure 5:** High-repetition rate 2DES setup. A beam splitter separates the NOPA beam into a pump and probe. Both, differential transmission and reflectivity spectra can be recorded on a single-shot basis

using a fast line camera and chopping. The other components are: variable neutral density filter (ND), iris aperture (IA), polarizer (POL), photo diode (PD), wedge pair (WP), vibrating mirror (VM), glass plate (GP) and focusing mirror (FM). WP: wedge pair, ND: neutral density filter, VM: vibrating mirror, POL: polarizer, OAP: off-axis parabolic mirror, PD: photo diode, GP: glass plate. A custom-made 500-slot chopper wheel is used to chop the pump at a rate of up to 50 kHz.

For pump-probe experiments, the coherence time is kept fixed at $\tau = 0$. For each waiting time the probe stage is moved and then a set number of spectra, e.g. 10000, are recorded. From these, 5000 differential spectra are calculated, averaged and saved. Pump-probe scans are then repeated a set number of times to increase signal-to noise ratio.

## 3. Two-dimensional electronic spectroscopy

In its simplest implementation, two-dimensional electronic spectroscopy (2DES) is a four-wave mixing technique that probes a third-order nonlinear material polarization which is induced by the interaction of the sample with two pump laser fields and one probe field.[42-45] Quite generally, this third order polarization $P^{(3)}(t)$ can be expressed in the time domain as a convolution of a third-order response function $R^{(3)}$ with the electric field profiles $E_i(t), i=1,2,3$ of the three laser pulses which arrive in the sample plane at times $t_{pu,1}, t_{pu,2}$ and $t_{pr}$, respectively (Fig. 6).[44-46]

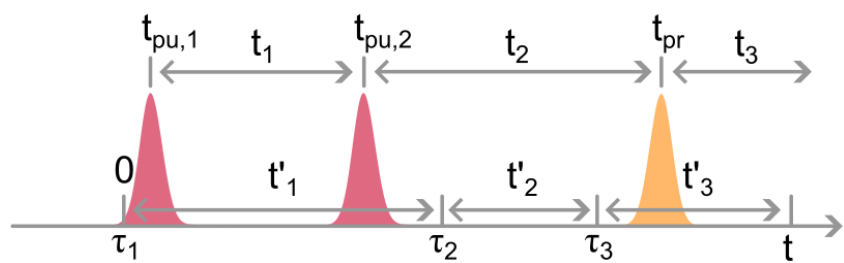

**Figure 6:** Time ordering of the laser field interactions in 2DES. The laser arriving in the sample plane at times $t_{pu,1}, t_{pu,2}$ and $t_{pr}$ and have electric field pulses $E_i(t'), i = 1,2,3$. The $\tau_i$ mark the absolute times of the light-matter interaction, while the $t'_i$ are relative times used in Eq. (2). The laser pulses with absolute times $t_{pu,1}, t_{pu,2}$ and $t_{pr}$ are separated by the relative times $t_i$.

When assuming strict time ordering, i.e., $t_{pr} > t_{pu,2} > t_{pu,1}$, and interactions of the three pulses with the sample at times $\tau_1 = 0, \tau_2 = t'_1$ and $\tau_3 = t'_1 + t'_2$, the third-order polarization (Fig. 6) can be expressed as

$$P^{(3)}(t) = \int_0^\infty dt'_3 \int_0^\infty dt'_2 \int_0^\infty dt'_1 \qquad (2)$$

$$E_3(t - t'_3) E_2(t' - t'_3 - t'_2) E_1(t' - t'_3 - t'_2 - t'_1) \cdot R^{(3)}(t'_3, t'_2, t'_1)$$

The third-order response function

$$R^{(3)}(t'_3, t'_2, t'_1) = \left(\frac{i}{\hbar}\right)^3 \left\langle \hat{\mu}_4 \left[\hat{\mu}_3 [\hat{\mu}_2, [\hat{\mu}_1, \hat{\rho}(-\infty)]]\right]\right\rangle \qquad (3)$$

describes the quantum dynamics of the matter system in response to delta-pulse excitations at the three interaction times. The interaction with the $i$-th field is described in point-dipole approximation using the dipole operator $\hat{\mu}_i = \hat{\mu}_I(\sum_{j=0}^{i-1} t'_j)$ with $t'_0 = 0$. In the interaction picture, the dipole operator $\hat{\mu}_I(t) = \exp\left(i\widehat{H_0}(t - t_0)\right) \hat{\mu} \exp\left(-i\widehat{H_0}(t - t_0)\right)$, where $\hat{\mu}$ is the electronic dipole moment operator in the Schrödinger picture and $\widehat{H_0}$ is the Hamiltonian of the system in the absence of a driving light field. The time $t_0$ denotes the start of the experiment. The fourth interaction with operator $\hat{\mu}_4$ is not related to any laser field interaction but evaluates $P^{(3)}(t)$ from the third-order perturbation of the density matrix of system $\hat{\rho}^{(3)}(t)$ at time $t$, $P^{(3)}(t) = \langle \text{Tr}\, \hat{\mu}_4 \hat{\rho}^{(3)}(t)\rangle$. Here, $\langle ... \rangle$ denotes the trace of the is matrix in brackets.

The response $R^{(3)}(t'_3, t'_2, t'_1)$ contains $2^n = 8$ terms, they are

$$R_1 = -\frac{i}{\hbar^3} \langle \hat{\mu}_4 \hat{\mu}_3 \hat{\mu}_2 \hat{\mu}_1 \hat{\rho}_0 \rangle = -\frac{i}{\hbar^3} \langle \hat{\mu}_4 \hat{\rho}_0 \hat{\mu}_1 \hat{\mu}_2 \hat{\mu}_3 \rangle^* = R_5^* \qquad (4)$$

$$R_2 = \frac{i}{\hbar^3} \langle \hat{\mu}_4 \hat{\mu}_3 \hat{\mu}_1 \hat{\rho}_0 \hat{\mu}_2 \rangle = \frac{i}{\hbar^3} \langle \hat{\mu}_4 \hat{\mu}_2 \hat{\rho}_0 \hat{\mu}_1 \hat{\mu}_3 \rangle^* = R_6^*$$

$$R_3 = \frac{i}{\hbar^3} \langle \hat{\mu}_4 \hat{\mu}_2 \hat{\mu}_1 \hat{\rho}_0 \hat{\mu}_3 \rangle = \frac{i}{\hbar^3} \langle \hat{\mu}_4 \hat{\mu}_3 \hat{\rho}_0 \hat{\mu}_1 \hat{\mu}_2 \rangle^* = R_7^*$$

$$R_4 = -\frac{i}{\hbar^3} \langle \hat{\mu}_4 \hat{\mu}_1 \hat{\rho}_0 \hat{\mu}_2 \hat{\mu}_3 \rangle = -\frac{i}{\hbar^3} \langle \hat{\mu}_4 \hat{\mu}_3 \hat{\mu}_2 \hat{\rho}_0 \hat{\mu}_1 \rangle^* = R_8^*.$$

Half of these terms ($R_1$ to $R_4$) are related to the remaining ones ($R_5$ to $R_8$) by complex conjugation. These terms are convoluted with the three laser fields. In general, this results in $(2 \cdot 3)^3 = 216$ field combinations. Imposing strict time-ordering, as done in Eq. (2), reduces the number of field interactions to $2^3 = 8$, therefore giving 64 terms in total (including those related by complex conjugation). These 64 terms can further be reduced by applying the rotating wave approximation and also by discarding those field interactions that do not result in a population state.[44-46] This gives a total of 16 response functions which are often displayed diagrammatically in the form of double-sided Feynman diagrams to find all contributions that remain. This diagrams are shown in Fig. 7 for an electronic three-level system comprising a ground state $|0\rangle$ (Energy $\hbar\omega_0$), an excited excited state $|1\rangle$ (Energy $\hbar\omega_1$) and a higher lying state $|2\rangle$ (Energy $\hbar\omega_2$). The laser field couples ground and excited state via a transition dipole moment $\mu_{01}$, and states $|1\rangle$ and $|2\rangle$ via $\mu_{12}$. We assume that the first pump pulse arrrives at time zero while the second pump is delayed by $\tau$. The probe pulse is time-delayed relative to the second pump by $T$.

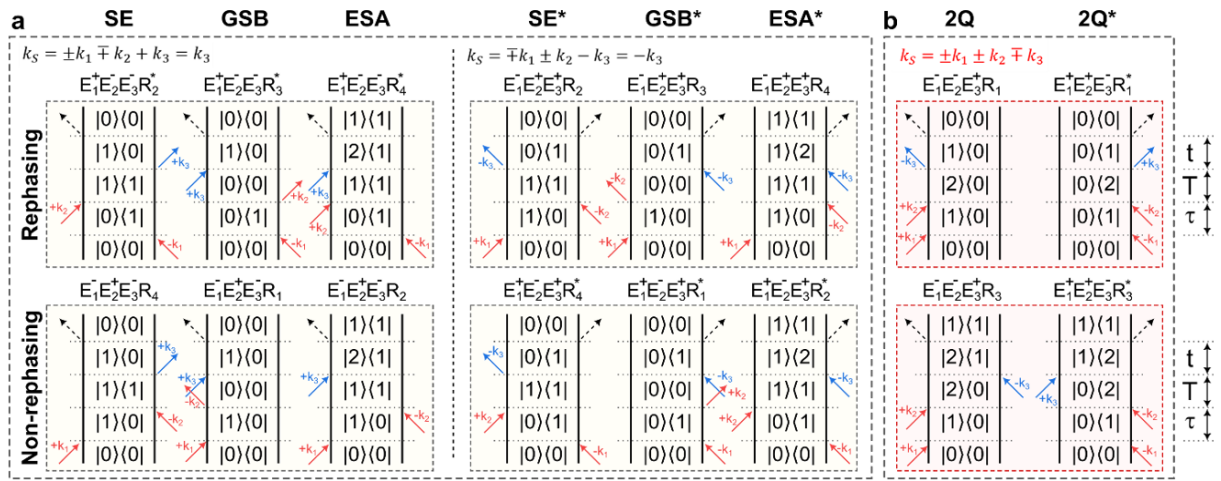

**Figure 2:** All double-sided Feynman diagrams for an electronic three-level system interacting with a well-ordered sequence of pulses (pump 1, pump 2, probe) that survive the rotating wave approximation and are resulting in a population state. **a:** *left:* Rephasing and non-rephasing diagrams representing stimulated emission (SE), ground-state bleaching (GSB) and excited state absorption (ESA) that are emitted into the $k_S = \pm k_1 \mp k_2 + k_3 = k_3$ direction. *right:* Set of complex conjugate pathways with conjugate wavevector $k_S = \mp k_1 \pm k_2 - k_3 = -k_3$. Since they do not provide additional information, they are usually not listed. In sum these 12 diagrams create the real-valued third-order response. **b:** Additional 2Q pathways that are not radiated into the phase-matched probe direction for this pulse ordering.

For more complex quantum systems, the three electronic states should be understood as manifolds with certain (for example vibronic) substructure. The system response is then obtained by performing three-dimensional summations over all those substates. Also in such a more general case, is it helpful to subdivide the system into a (i) ground state manifold, (ii) a manifold of single-quantum states, coupled to the ground state and by an optical transition dipole moment and (iii) a manifold of two-quantum states which are decoupled from the ground states but can be populated by an optical transition from the one-quantum states. These 16 diagrams have a clear and transparent physical meaning which is schematically depicted in Fig. 8. The first pump pulse creates an electronic coherence between states $|0\rangle$ and $|1\rangle$ which decays with the decoherence time $T2_{01} = 1/\gamma$ for the $|0\rangle \to |1\rangle$ transition. If the second pump arrives while the coherence is still present, it transforms the coherence into a population in either $|0\rangle$ and $|1\rangle$ state. Importantly, the amplitude of this population depends sensitively on the phase of the electronic coherence, allowing to periodically modulate the population by tuning $\tau$ with sub-cycle resolution[47]. In the more general case of ground-state and single-quantum state manifolds, the second pump will launch coherent wavepackets in the respective states. During the waiting time T, the populations (the ground and excited state wavepackets) will then evolve in time.[48, 49] The resulting quantum dynamics are then interrogated by the probe laser, creating a third-order electronic coherence between either states $|0\rangle$ and $|1\rangle$ or states $|1\rangle$ and $|2\rangle$. The electric field that is emitted from the corresponding non-linear polarization is detected in the experiment.

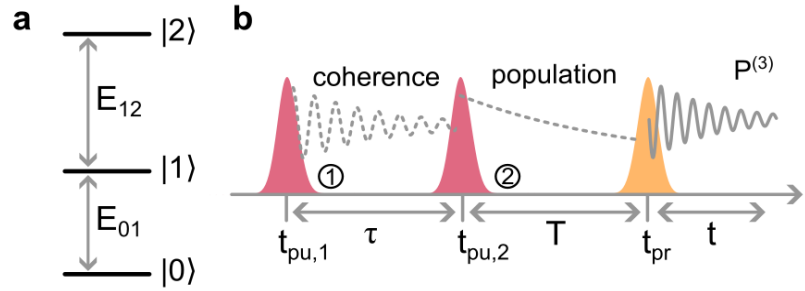

**Figure 8:** Scheme of the interaction of an electronic three-level system with a sequence of three well-ordered pulses with $T > 0$ and $\tau > 0$. Within third-order perturbation theory, the quantum pathway that leads to a 2DES signal, consists of three steps. First, the first pump pulse creates a coherence in the density matrix between states $|0\rangle$ and $|1\rangle$ that evolves during the dephasing time. Then, the second pump pulse converts this coherence into a population in either states $|0\rangle$ or $|1\rangle$. During the dephasing time, the amplitude of the population depends sensitively on the phase of the coherence. During $T$, the population evolves until it is probed by the third pulse. The probe creates a second coherence, which corresponds to the third-order polarization that is detected in a 2DES experiment.

Consequently, the Feynman diagrams are usually grouped into ground state bleaching (GSB) diagrams, probing population (wavepacket) dynamics in the electronic ground state and stimulated emission (SE) and excited state absorption (ESA) diagrams, which are governed by the population (wavepacket) dynamics in the excited state. The excited state dynamics can either be read out via probe-induced stimulated emission from $|1\rangle$ to $|0\rangle$ (SE) or via an excited state absorption from $|1\rangle$ to a higher-lying state $|2\rangle$. Addditionally, the Feynman diagrams are classified into "rephasing" (R) and "non-rephasing" (NR) diagrams. For the rephasing diagrams, the coherences induced by the first pump and the probe oscillate at frequencies of opposite sign. For the non-rephasing diagrams, instead, both coherences are either oscillating at positive or negative frequencies with the same sign. In inhomogeneously broadened systems, the rephasing diagrams can generate photon echos[50, 51] in the nonlinear response and allow for separating homogeneous and inhomogeneous broadening of the system[52]. When considering plane wave excitation of the sample by a collinearly propagating pair of pump pulses, as in the present implementation of 2DES, the electric fields emitted by all GSB, SE and ESA diagrams propagate in the direction of the probe laser. R and NR can be separated by choosing a noncollinarly propagating pump pair, e.g, in a box car geometry[53] or by applying certain phase cycling schemes.[54, 55]

So far, we have discussed those Feynman diagrams that create populations after the interaction with the two pulses. In addition, the electronic coherence induced by the first pump pulse may also be transferred into an electronic coherence between $|0\rangle$ and $|2\rangle$, providing direct access to the (manifold of) two-quantum states. In third-order perturbation, the dephasing of this coherence is sensed via the two-quantum (2Q) Feynman diagrams depicted in Fig. 7. For plane-wave excitation, the 2Q signals will be emitted in the direction $k_{2Q} = +k_1 + k_2 - k_3$. Therefore, they will not be emitted in the probe direction for the excitation scenario pump 1 – pump 2 – probe considered so far. They are made accessible by altering the pulse ordering.[56, 57]

To give analytical expressions for the response functions that result from these Feynman diagrams, we assume an identical dephasing rate $\gamma$ for the $|0\rangle \rightarrow |1\rangle$ and for the $|1\rangle \rightarrow |2\rangle$ transition and use a 2Q-dephasing rate of $\gamma_{2Q}$ for the $|0\rangle \rightarrow |2\rangle$ transition. In addition, we consider an incoherent population relaxation of the $|1\rangle$ state at rate $\kappa$. We give the response functions for one half of the 16 diagrams. The other half can be obtaind by complex conjugation. The different responses are the **rephasing**

$$R_R^{SE}(t,T,\tau) = -\frac{i}{\hbar^3}\mu_{01}^4 e^{+\frac{i}{\hbar}(E_1-E_0)\tau-\left(\gamma+\frac{\kappa}{2}\right)\tau} e^{-\frac{i}{\hbar}(E_1-E_1)T-\kappa T} e^{-\frac{i}{\hbar}(E_1-E_0)t-\left(\gamma+\frac{\kappa}{2}\right)t} \quad (5)$$

$$R_R^{GSB}(t,T,\tau) = -\frac{i}{\hbar^3}\mu_{01}^4 e^{+\frac{i}{\hbar}(E_1-E_0)\tau-\left(\gamma+\frac{\kappa}{2}\right)\tau} e^{-\frac{i}{\hbar}(E_0-E_0)T-\kappa T} e^{-\frac{i}{\hbar}(E_1-E_0)t-\left(\gamma+\frac{\kappa}{2}\right)t}$$

$$R_R^{ESA}(t,T,\tau) = +\frac{i}{\hbar^3}\mu_{01}^2\mu_{12}^2 e^{+\frac{i}{\hbar}(E_1-E_0)\tau-\left(\gamma+\frac{\kappa}{2}\right)\tau} e^{-\frac{i}{\hbar}(E_1-E_1)T-\kappa T} e^{+\frac{i}{\hbar}(E_2-E_1)\tau-\left(\gamma+\frac{\kappa}{2}\right)\tau}$$

**non-rephasing**

$$R_{NR}^{SE}(t,T,\tau) = -\frac{i}{\hbar^3}\mu_{01}^4 e^{-\frac{i}{\hbar}(E_1-E_0)\tau-\left(\gamma+\frac{\kappa}{2}\right)\tau} e^{-\frac{i}{\hbar}(E_1-E_1)T-\kappa T} e^{-\frac{i}{\hbar}(E_1-E_0)t-\left(\gamma+\frac{\kappa}{2}\right)t} \quad (6)$$

$$R_{NR}^{GSB}(t,T,\tau) = -\frac{i}{\hbar^3}\mu_{01}^4 e^{-\frac{i}{\hbar}(E_1-E_0)\tau-\left(\gamma+\frac{\kappa}{2}\right)\tau} e^{-\frac{i}{\hbar}(E_0-E_0)T-\kappa T} e^{-\frac{i}{\hbar}(E_1-E_0)t-\left(\gamma+\frac{\kappa}{2}\right)t}$$

$$R_{NR}^{ESA}(t,T,\tau) = \frac{i}{\hbar^3}\mu_{01}^2\mu_{12}^2 e^{-\frac{i}{\hbar}(E_1-E_0)\tau-\left(\gamma+\frac{\kappa}{2}\right)\tau} e^{-\frac{i}{\hbar}(E_1-E_1)T-\kappa T} e^{-\frac{i}{\hbar}(E_2-E_1)t-\left(\gamma+\frac{\kappa}{2}\right)t}$$

and **2Q pathways**

$$R_1^{2Q}(t,T,\tau) = -\frac{i}{\hbar^3}\mu_{01}^2\mu_{12}^2 e^{-\frac{i}{\hbar}(E_1-E_0)\tau-\left(\gamma+\frac{\kappa}{2}\right)\tau} e^{-\frac{i}{\hbar}(E_2-E_0)T-\gamma_{2Q}T} e^{-\frac{i}{\hbar}(E_1-E_0)t-\left(\gamma+\frac{\kappa}{2}\right)t} \quad (7)$$

$$R_2^{2Q}(t,T,\tau) = \frac{i}{\hbar^3}\mu_{01}^2\mu_{12}^2 e^{-\frac{i}{\hbar}(E_1-E_0)\tau-\left(\gamma+\frac{\kappa}{2}\right)\tau} e^{-\frac{i}{\hbar}(E_2-E_0)T-\gamma_{2Q}T} e^{-\frac{i}{\hbar}(E_2-E_1)t-\left(\gamma+\frac{\kappa}{2}\right)t}.$$

Experimentally, we measure the electric field that is emitted by the nonlinear polarization. For a spatially homogeneous sample, this real-valued field phase lags the polarization by $\pi/2$ [44, 58] and is given as

$$S_{2D}(t,T,\tau) = \sum_j iR_j(t,T,\tau) - iR_j^*(t,T,\tau). \quad (8)$$

The spectrometer then measures the spectrum the arises from the heterodyne measurement with the probe. It can be expressed as a Fourier transform along t

$$S_{2D}(\omega_{det},T,\tau) = -\Re\left\{\int_{-\infty}^{\infty}\Theta(t)\,S_{2D}(t,T,\tau)e^{+i\omega_{det}t}\,dt\right\}, \quad (9)$$

where the minus sign accounts for the differential transmission, $\Delta T/T$, measurement that is actually performed and the real part is taken since the signal is measured at positive frequencies $\omega_{det}$ using a spectrometer. $\Theta(t)$ denotes the Heaviside step function. To obtain the 2DES signals, this spectrogram is numerically Fourier transformed along the coherence time axis $\tau$

$$S_{2D}(\omega_{det},T,\omega_{ex}) = \int_{-\infty}^{\infty}\Theta(\tau)\,S_{2D}(\omega_{det},T,\tau)e^{+i\omega_{ex}\tau}d\tau \quad (10)$$

to obtain a (complex-valued) energy-energy map as a function of the waiting time $T$. Usually, in our experiments, only the real part of these maps is recorded, measuring absorptive 2DES maps.

The signals that are observed in such 2DES map have a very transparent physical meaning, as can be seen by closer inspection of the Feynman diagrams in Fig. 7.[44, 45, 59] The excitation by the first pump pulse creates peaks along the excitation axis, $\omega_{ex}$, that are centered around all ground to excited state transitions at frequencies $\omega_{01} = (E_1 - E_0)/\hbar$ that are optically allowed ($\mu_{01} \neq 0$). Peaks with positive amplitude, induced by GSB and SE diagrams, appear along the detection axis due to the interaction of the probe with the system. New peaks with negative amplitude (induced absorption) appear at frequencies $\omega_{12} = (E_2 - E_1)/\hbar$ if the excitation by the pump induces new optical excitations in the system. Thus, a diagonal peak in the 2DES map implies that both first pump and probe interact with the same transition while an off-diagonal peak shows that the excitation of one resonance alters the interaction of the probe with a second resonance. In the case of a homogeneously broadened system, the lineshape of each peak is defined by the homogeneous dephasing times of involved coherences while inhomogeneous and homogenous broadening can be separated in the case of an inhomogeneously broadened system.[60] Of particular interest are the waiting time dynamics of the 2DES peaks. Quite generally, they probe the quantum dynamics of the wavepackets that are created in either the ground or the excited state by the interaction with the second pump laser. Two classes of dynamics should be distinguished. Incoherent energy transfer process between different quantum states usually result in gradual changes in peak amplitudes with $T$. In contrast, strong coupling between excited quantum states will result in the formation of new hybrid states $|i\rangle$ with frequencies $\omega_i$. The resulting 2DES maps show peaks at the frequencies $\omega_{0i}$. Importantly the amplitude of these peaks may periodically oscillate at the difference frequency between two of these hybrid states. Such "beating peaks" are generated if the two pump pulses interact with two different hybrid states. Such quantum beats are a distinct signature of strong couplings in the investigated quantum system. This qualitative discussion so far ignores phase-breaking decoherence (pure dephasing) and energy-relaxation processes.

### 3.1 Phase cycling with the TWINS interferometer

It is evident from the Feynman diagrams in Fig. 7 that a separation of rephasing and non-rephasing diagrams is needed to isolate certain response functions and – thus – certain pathways in the light-matter interaction. Such a separation can be used, e.g., to distinguish between ground and excited state vibrational wavepacket motion in molecular systems[49] or to disentangle many-body interactions in semiconductors.[61, 62] Generally, this requires some degree of freedom in the field interaction is required. For example, directional phase matching can be used in the BoxCARS geometry, taking advantage of the different angles under which the three pulses impinge on the sample.[63, 64] For partially (or fully) collinear 2DES, careful manipulation of the phases of the pulses via phase-cycling (PC)[54, 55, 65-67] or frequency-tagging[68-70] is required. Here, 2DES data are recorded for different absolute phase values that are imprinted onto the excitation pulses and the desired quantum pathways are then isolated from a linear combination of the nonlinear signals.

In addition to isolating R and NR pathways, PC also allows to record a distinctly different class of 2DES experiments[63, 71, 72] termed as zero-quantum (0Q) and double-quantum (2Q) 2DES. In contrast to 1Q 2DES, 0Q and 2Q 2DES probe coherences between states that are separated by zero and two quanta of the excitation energy, respectively. They therefore provide information about two-quantum coherences and coherences within the excited or ground state manifold. This can give additional and direct insights into coherent couplings, (many-body) interactions and doubly-excited states in the system.[72-75]

Until recently, we (and others[76]) believed that it is difficult to implement such phase-cycling schemes in the conceptually simple pump-probe geometry afforded by the TWINS. We therefore thought that the use of the TWINS prohibits isolation of R and NR pathways and recording of 0Q and 2Q spectra. Very recently, we reported a simple and straightforward solution to the phase-cycling problem for TWINS.[36] We showed that the insertion of an achromatic quarter-wave plate (QWP) (Fig. 9) allows to manipulate the two polarization components inside the TWINS and unlocks full control of the relative absolute phase between the pump pulses. This opens up advanced 2DES schemes for TWINS such as isolating R and NR contributions but also 0Q and 2Q spectroscopy. In Section 5, some of these new capabilities are demonstrated for 2DES on a J-aggregated squaraine thin-film at room temperature.

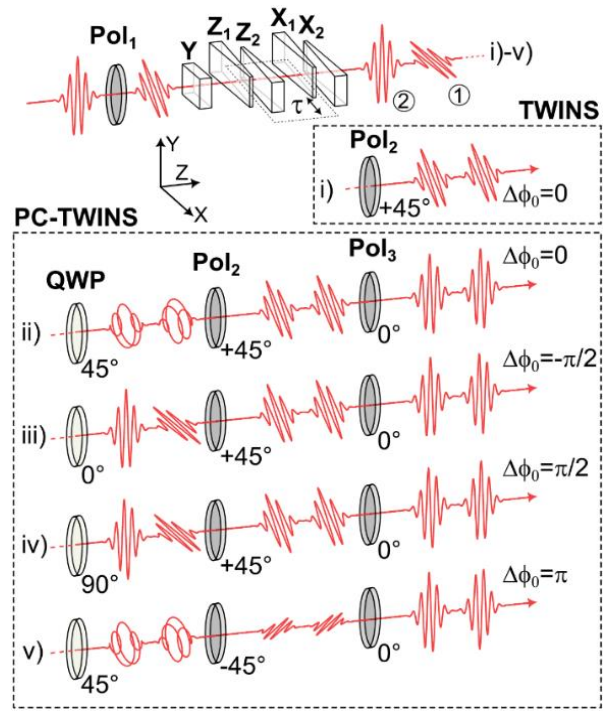

**Figure 9:** Scheme of the PC-TWINS for phase-cycling. The next row, case i), depicts the traditional TWINS case where a second polarizer projects both pulses onto a common polarization direction. For the PC-TWINS a quarter-wave plate (QWP) introduces a spectrally-constant phase-difference $\Delta\phi_0$ between the two output pulses. ii)-iv): Orientations of the fast axis at 45°, 0° and 90° introduce phase differences of $0, -\pi/2$ and $+\pi/2$, respectively. The second polarizer (Pol$_2$) projects the polarization of both pulses onto 45°. v): When Pol$_2$ is turned from +45° to -45°, an additional phase of $\pi$ is introduced. The final and optional polarizer (Pol$_3$) ensures constant polarization and output intensity for all displayed phase settings. These figures are reprinted with permission from Ref. [36], Copyright (c) 2024 Optica Publishing Group.

## 4. Vibronic couplings in squaraine dyes in solution

In this and the following Sections, we will mainly focus on the optical properties of a prototypical quadrupolar D-π-A-π-D molecule, 2,4-Bis[4-((S))-2-(hexadecyloxymethyl)-pyrrolidone-2,6-dihydroxyphenyl] squaraine molecules (ProSQ-C16 or in short squaraine), Fig. 1a.

(S,S)-enantiomers of the squaraine have been synthesized as described by Schulz et al. [77]. The molecule consists of a squaric acid acceptor at the center of the quadrupolar D-π-A-π-D molecule with two aniline donor arms on the left- and right-hand sides.[15, 78] The four hydroxyl groups at the phenyl groups connecting donor and acceptor stabilize the molecule in a straight planar geometry by forming hydrogen bonds to the squaric acid's carbonyl groups and do not allow for any bending angle between the two arms. The side chains, which have a minor effect on the electronic structure and mainly influence the aggregation properties, point out of the molecular plane. Two different enantiomers (R,R and S,S) with side chains pointing in an either clockwise or counter-clockwise direction, dictate the circular dichroism of the ProSQ-C16 molecular aggregates.[15] In the experiments reported in this Section we focus on squaraines dissolved in chloroform liquid.

A conceptually appealing essential state model (ESM) to intuitively account for the most salient electronic and optical properties of such quadrupolar dyes has been introduced by Painelli and coworkers.[2, 16-18] It considers the molecules as being comprised of two polar dyes in either a neutral ground $|DA\rangle$ or charge-separated excited state $|D^+A^-\rangle$, connected via a charge-accepting linker group. Consequently, three essential electronic states are identified as a minimum (diabatic) basis for quadrupolar dyes: a neutral state $|N\rangle = |DAD\rangle$ and two charge-separated states $|Z_1\rangle = |D^+A^-D\rangle$ and $|Z_2\rangle = |DA^-D^+\rangle$. The latter two are degenerate zwitterionic states having ionic charge distributions on either the left or the right arm of the molecule. The neutral state is separated by an energy gap $\eta_z$ from the zwitterionic states which are coupled to $|N\rangle$ via a charge-transfer integral $t_z$. Both $\eta_z$ and $t_z$ are free parameters of the ESM Hamiltonian. The diagonalization of this Hamiltonian yields three "adiabatic" electronic states as linear combinations of the diabatic states: a ground state $|g\rangle$ with partial charge transfer (CT) character $D^{+\rho/2}A^{-\rho}D^{+\rho/2}$, a first excited state $|c\rangle$ ($D^{+1/2}A^{-1}D^{+1/2}$) and a second excited state $|e\rangle$

($D^{+(1-\rho)/2}A^{-(1-\rho)}D^{+(1-\rho)/2}$). Here, the parameter $\rho$ expresses the contribution of the zwitterionic states to $|g\rangle$ and thus the strength of the CT character of the ground state. Importantly, the optical transition between $|g\rangle$ and $|c\rangle$ is one-photon allowed, whereas the second excited state, separated from $|c\rangle$ by an energy gap that is set by the strength of the CT integral,[2] can only be accessed by two-photon transitions.[9, 79-83]

A second, important prediction of the ESM is a distinct modification of vibronic couplings in quadrupolar dyes.[16, 17] In polar organic dyes, electronic excitations are typically strongly coupled to high-frequency stretching modes of the carbon backbone, resulting in substantial displacements, $\Delta_{dia}$, on the order of unity, of the excited state potential energy surface along the dimensionless coordinate of the stretching mode.[84, 85] In quadrupolar squaraine dyes, vibronic coupling results in a coupling of all three adiabatic states to the Raman-active, symmetric superposition $|+\rangle$ and the infrared-active antisymmetric superposition $|-\rangle$ of the high-frequency stretching modes on either arm of the molecule.[2, 16, 83] Effectively, this is expected to result in a substantial reduction of the relative displacement $\Delta_{gc} = (1-\rho)\Delta_{dia}/\sqrt{2}$ between the ground and one-photon allowed $|c\rangle$ state, compared to $\Delta_{dia}$ of the constituent polar dye. This leads to a much smaller Huang-Rhys factor $S_{gc} = \Delta_{gc}^2/2$ that characterizes the coupling to the symmetric stretching vibrations. So far, this implication of the ESM for the vibronic coupling has mainly been investigated theoretically.

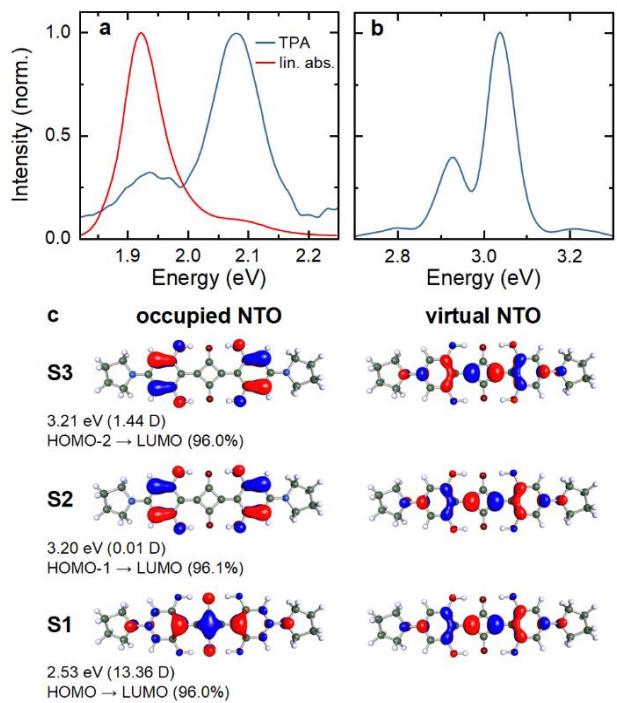

**Figure 10:** Electronic structure of the squaraine dye ProSQ-C16. **a**: Two-photon absorption (TPA) of the optically allowed $|g\rangle \rightarrow |c\rangle$ transition in chloroform (blue), compared to the linear absorption spectrum (red). **b**: TPA in the region of the one-photon forbidden $|g\rangle \rightarrow |e\rangle$ transition at 3.04 eV. **c**: Calculated excited states S1-S3 in vacuum with their excited state energy and transition dipole moment in Debye (D). Orbital plots show virtual and occupied natural transition orbitals (NTOs) of the electron (right) and hole (left) pair associated with the S1-S3 states. All calculated NTOs are dominated by a single transition between molecular orbitals as indicated. These figures are reprinted with permission from Ref. [19], Copyright(c) 2016 APS Journals.

### 4.1. Electronic structure of squaraine molecules

To gain insight into energetic ordering of the electronic states of the squaraine molecule and to probe the optical properties of the one-photon forbidden $|e\rangle$ state, predicted by the ESM, two-photon absorption spectra (TPA) have been recorded. TPA in the $|g\rangle \rightarrow |c\rangle$ region (Fig. 10a, blue line) shows a strong peak around 2.1 eV, the location of the weak vibrational shoulder in linear absorption. In contrast, TPA around the main OPA resonance (red line) at 1.93 eV is much reduced. The splitting between the two TPA peaks is 160 meV (1300 cm$^{-1}$), approximately given by the energy of high-frequency carbon backbone mode considered in ESM[9, 82, 86, 87] In Fig. 10b, we probe the spectral region around 3 eV, predicted by the ESM to contain the two-photon allowed $|g\rangle \rightarrow |e\rangle$ transition.[2, 9, 10, 80, 82, 86] Indeed, our measurement shows the expected pronounced TPA signal around 3.04 eV. We also observe a clear substructure of this peak with a strong shoulder at 2.93 eV. No noticeable OPA is seen in this region, pointing to a planar symmetric molecular structure of the molecule without bending of the two arms.[88]

The findings can be rationalized on the basis of the ESM using second order perturbation theory. Three "adiabatic" $|g\rangle$, $|c\rangle$ and $|e\rangle$ states are coupled to the Raman-active, symmetric $|+\rangle$ and infrared-active, anti-symmetric $|-\rangle$ vibrational modes that result from the hybridization of the high-frequency carbon backbone vibrations on the left and right arm of squaraine molecule. Only the $|c\rangle$ state can be reached from the ground state by a one-photon transition. Consequently, TPA into the manifold of vibrational levels in $|c\rangle$ can only be achieved by combining a one-photon, dipole-allowed $|g\rangle \rightarrow |c\rangle$ transition with a highly off-resonant excitation of the infra-red active $|-\rangle$ mode by the laser field. Thus, the lowest energy TPA-allowed transition to $|c\rangle$ is expected at an energy $E_{gc} + \hbar\omega_-$, where $E_{gc} = E_c - E_g$ ($E_i$ being the energy of adiabatic state $|i\rangle$) and $\omega_-$ is the frequency of the anti-symmetric $|-\rangle$ mode. This suggests a dominant antisymmetric vibrational mode energy of $\hbar\omega_- = 160$ meV. In this picture, two-photon allowed peaks around 3 eV are dominated by a matrix element associated with sequential dipole-allowed $|g\rangle \rightarrow |c\rangle$ and $|c\rangle \rightarrow |e\rangle$ transitions. Since the intensity of vibrational sidebands in these transitions is weak, the dominant TPA peak in Fig. 10b is expected to reflect the purely electronic transition between $|g\rangle$ and $|e\rangle$ with $E_{ge} = E_e - E_g = 3.04$ eV.

Based on this analysis, we can now determine the ESM parameters for the investigated ProSQ-C16 molecule. The values for $\eta_z$ and $t_z$ can be calculated from the adiabatic eigen energies[16]

$$E_g = \frac{1}{2}\left(\eta_z - \sqrt{\eta_z^2 + 8t_z^2}\right) \qquad (11)$$
$$E_c = \eta_z$$
$$E_e = \frac{1}{2}\left(\eta_z + \sqrt{\eta_z^2 + 8t_z^2}\right).$$

With $E_{gc} = 1.93$ eV and $E_{ge} = 3.04$ eV we obtain $\eta_z = 0.8$ eV and $t_z = 1.04$ eV. Using these values for the zwitterionic energy gap and the CT integral we obtain the $\rho$ parameter from

$$\rho = \frac{1}{2}\left(1 - \frac{\eta_z}{\sqrt{\eta_z^2 + 8t_z^2}}\right) \quad (12)$$

as $\rho = 0.37$, which is a reasonable value for squaraines and confirms the assignment as a type II D-A-D chromophore within the ESM.[2] The value of $\rho$ fixes the amount of CT in the ground state $|g\rangle$ $(D^{+\rho/2}A^{-\rho}D^{+\rho/2})$ and in the second excited, dark state $|e\rangle$ ($D^{+(1-\rho)/2}A^{-(1-\rho)}D^{+(1-\rho)/2}$), whereas the CT character of the optically bright state $|c\rangle$ ($D^{+1/2}A^{-1}D^{+1/2}$) is $\rho$-independent.

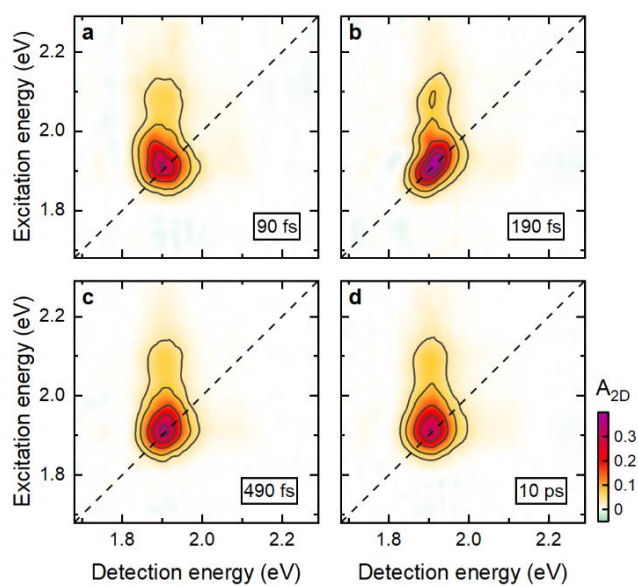

**Figure 11.** 2DES maps of the squaraine moleculeProSQ-C16 in chloroform for selected waiting times. Vibrational substructure with ~180 meV, corresponding to the fast ~22 fs oscillations observed in pump-probe dynamics, show up along the excitation and detection at ~2.1 eV. Spectral signatures arising from the 147 cm$^{-1}$ mode (18 meV) and the 570 cm$^{-1}$ mode (~71 meV) can be seen as a periodic distortion of the diagonal peak, especially noticeable at (**a**) 90 and (**b**) 190 fs. These figures are reprinted with permission from Ref. [19], Copyright (c) 2016 APS Journals.

Quantum chemical calculations support these conclusions. The natural transition orbitals[89] (NTOs) for the lowest three excited states, S1-S3, are calculated to give an intuitive picture of the optical excitations in the molecule. As depicted in Fig. 10c, the occupied and virtual NTOs of the S1 - S3 states directly visualize the electron and hole orbitals associated with the optical excitation of the molecule, respectively.[80] The NTOs of S1-S3 are dominated (~96%) by transitions between single molecular orbitals, namely S1: HOMO → LUMO, S2: HOMO-

1→LUMO and S3: HOMO-2→LUMO with transition energies of 2.53 eV, 3.20 eV and 3.21 eV, respectively. Among these lowest excited states, only S1 has a substantial transition dipole moment of $\mu_{S1} = 13.36$ D, oriented along the long axis of the molecular plane. We can thus associate S1 with the $|c\rangle$ state in the ESM and $\mu_{S1}$ with the transition dipole moment for the optically bright $|g\rangle \rightarrow |c\rangle$ transition. The value for $\mu_{S1}$ predicted by theory is in excellent agreement with the experimentally obtained value of $12.5 \pm 0.6$ D. The simulations predict that S2 is optically dark, $\mu_{S2} = 0.01$ D, and we thus assign it to the $|e\rangle$ state of the ESM. S3 has a weak transition dipole moment of $\mu_{S3} = 1.44$ D, that lies within the molecular plane and is oriented along its short axis. This state is not covered by the simplified three-state ESM.[79]

### 4.2. 2D electronic spectroscopy of squaraine molecules

To further support the conclusions drawn about the electronic structure of the squaraine molecule, we performed 2DES experiments of the molecule in solution, giving access to excitation and detection energy dependent differential spectra as a function of waiting time. Such 2DES maps are displayed in Fig. 12 for four selected waiting times. The 2DES maps are dominated by a diagonal peak at ~1.9 eV (the central absorption peak of the $|g\rangle \rightarrow |c\rangle$ transition) with additional cross peaks at ~2.1 eV along both the excitation and detection axes. Since the signal amplitude of the 2DES maps is positive, $A_{2D} > 0$, the signal is governed by GSB of the $|g\rangle \rightarrow |c\rangle$ transition and SE from $|c\rangle \rightarrow |e\rangle$, while excited state absorption contributions are much weaker. The effect of the strong low-frequency modes on the 2DES spectra manifests itself again as a distortion of the diagonal peak, especially well noticeable in Fig. 4a-b. Since the energy corresponding to the high-frequency ~22 fs vibrations (~180 meV) is larger than the linewidth, the resulting DHO peak structure is well resolved in the 2DES spectra. Coupling to lower frequency vibrations results in periodically oscillating distortions of the waiting-time dependent 2DES line shape (Fig. 11), which are more challenging to analyze quantitatively since several modes contribute to those distortions.

## 4.3. Vibronic couplings in squaraine molecules in solution: Experiment

To investigate vibronic couplings in squaraine molecules, we employ time-resolved vibrational spectroscopy using an ultrashort, sub-10 fs pump pulse, tuned in resonance with the $|g\rangle \rightarrow |c\rangle$ transition. The optical resonant excitation creates a coherent vibrational wavepacket in the Franck-Condon region of the excited state potential energy surface (PES). If the minimum of the excited state PES is shifted by a dimensionless displacement $\Delta_i$ with respect to the ground state PES along the dimensionless vibrational coordinate $Q_i$, the optical excitation triggers coherent wavepacket oscillations around the excited state PES minimum.[90] This oscillation, with period $\tau_{vib,i} = 2\pi/\omega_{vib,i}$ ($\omega_{vib,i}$ frequency of the vibrational mode i), will then periodically modulate the transmission spectrum of a second ultrashort probe pulse (displacive excitation in the displaced harmonic oscillator (DHO) model).[48, 90] The differential transmission spectrum $\Delta T/T(T, E_{det})$ as a function of waiting time T and detection energy $E_{det}$ directly tracks the coherent wavepacket motion in the time domain.

Fig. 12a depicts the result of such a pump-probe measurement performed on squaraine molecules in chloroform at room temperature after removal of coherent solvent vibrations and cross-phase modulation signals.[91] For positive waiting times, the main feature in the map in Fig. 12a is a positive SE and GSB band at the position of the $|g\rangle \rightarrow |c\rangle$ transition at ~1.9 eV with weak shoulders at ~1.75 eV and ~2.1 eV (Fig. 12e). At negative waiting times, we observe signatures arising from the perturbed free induction decay[92] of the $|g\rangle \rightarrow |c\rangle$ transition. $\Delta T/T$ shows persistent periodic modulations with periods of $\tau_{vib,1} = 227$ fs (147 cm$^{-1}$) and $\tau_{vib,2} = 58$ fs (570 cm$^{-1}$) (Fig. 12f). Additional faint high-frequency vibrational oscillations with $\tau_{vib,3} \approx 22$ fs (1500 cm$^{-1}$) are only resolved on the high-energy side of the SE/GSB peak, as highlighted in the inset of Fig. 12a. To isolate the oscillatory part of the signal, we subtract a constant background for waiting times beyond 50 fs (Fig. 12b). A Fourier transform of these residuals along the waiting time (Fig. 12c-d) confirms two dominant modes with 147 cm$^{-1}$ and

570 cm$^{-1}$. We estimate a vibrational relaxation time $T_{vib,1}$ of the low frequency mode of 350 fs, whereas that of the high-frequency 570 cm$^{-1}$ mode is much longer, $T_{vib,2} \approx 1.5$ ps. Weaker peaks at 367 cm$^{-1}$ and 668 cm$^{-1}$ in the Fourier spectra arise from remaining chloroform Raman signals.[93] Even though ~1500 cm$^{-1}$ modes appear as faint ~22 fs oscillations in the time domain, their amplitude is too weak to provide clear peaks in the Fourier spectra.

The experimental pulse duration of <10 fs is much shorter than the vibrational periods $\tau_{vib,1}$ and $\tau_{vib,2}$. In this limit, the oscillations in the pump-probe signal are dominated by excited state wavepacket motion whereas impulsive Raman excitation of ground state wavepackets is weak. The coherent vibrational dynamics in $\Delta T/T$ can be well described with the classical limit of the DHO.[48] Since the energy of both modes is smaller than the linewidth of the electronic $|g\rangle \rightarrow |c\rangle$ transition, the dominant effect of impulsive excitation is a waiting time-dependent modulation of the transition energy following

$$\Delta E_{gc}(T) = \sum_i \hbar\omega_{\text{vib},i}\Delta_i^2 \cos(\omega_{vib,i}T)\, e^{-T/T_{vib,i}}. \qquad (13)$$

This energy modulation can be understood as the continuous, time-dependent change in the difference between the ground and excited state PES traced by the center of mass of the coherently oscillating excited state wavepacket. Upon impulsive excitation, the wavepacket is launched in the Franck-Condon region of the excited state PES, with maximum energy difference $\Delta E_{gc}$ to the ground state. After half an oscillation period, the wavepacket reaches the outer turning point with a minimum energy in $\Delta E_{gc}$. The resulting periodic modulation in transition energy introduces a spectral modulation pattern of the differential transmission with an enhanced amplitude on the sides of the resonance and a π phase flip in the center. The displacement between ground and excited state PES dictates the strength of these amplitude oscillations. As such, Eq. (5) allows us to directly estimate the displacements for the low-frequency modes from the experimental data as $\Delta_1 \approx 0.8$ and $\Delta_2 \approx 0.2$.

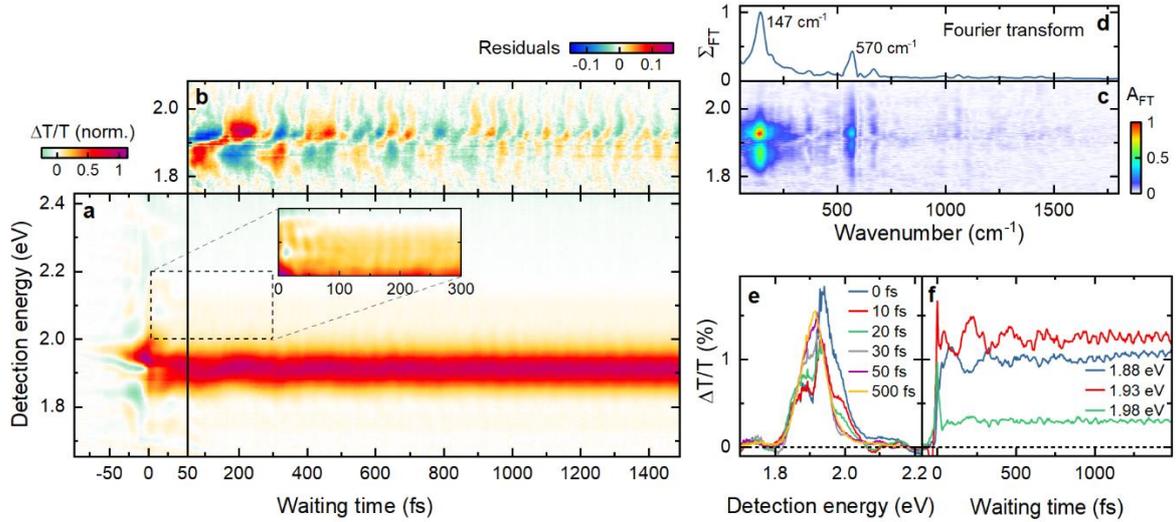

**Figure 12:** Pump-probe measurements of the squaraine molecule in chloroform revealing coherent vibrational wavepacket motion. **a:** Solvent corrected pump-probe map, showing a strong bleaching and stimulated emission band at ~1.9 eV. At positive waiting times, the signal undergoes periodic oscillations with multiple frequencies. At ~2.1 eV faint ~22 fs oscillations can be observed (inset). **b:** Residuals reveal 227 fs and 58 fs-period oscillations. **c:** Fourier transform of the residuals in **b**. **d:** Vibrational spectrum obtained by summing up the map in **c**, showing two dominant peaks at 147 cm$^{-1}$ and 570 cm$^{-1}$. **e:** Spectral evolution of the differential signal for selected waiting times. **f:** Waiting time dynamics for three selected detection energies. These figures are reprinted with permission from Ref. [20], Copyright (c) 2016 APS Journals.

Couplings to high-frequency modes with energies that are larger than the electronic linewidth cannot be described in this classical limit. For such modes, the sidebands that result from the coherent vibrational wavepacket motion, are well resolved in the optical spectra. Here, an upper limit for the strength of the vibronic coupling is given by the amplitudes of the *n*-th vibrational sideband, which follow a Poisson distribution $I(n) = I(0)S^n/n!$.[16] The peak ratio between the first vibrational sideband and the zero phonon line in the absorption spectrum in Fig. 1 sets an upper limit of $\Delta_3 \approx 0.45$ for the effective displacement of the ~1500 cm$^{-1}$ modes in chloroform. This value, however, does not agree with the results of our pump-probe measurement as it predicts much larger oscillation amplitudes than observed experimentally. From the very faint fast oscillations in Fig. 3a, we estimate an upper limit of $\Delta_3 \approx 0.25$, corresponding to a Huang-Rhys factor $S_3 \lesssim 0.03$.[16] These pump-probe measurements therefore conclusively show that the electronic excitations of the squaraine molecule are weakly coupled to high-frequency ~1500 cm$^{-1}$ modes and that modulations of its optical response are dominated by couplings to low-frequency vibrations.

### 4.4. Vibronic couplings in squaraine molecules in solution: Theory

We now compare the experimental results to a theoretical modeling of the vibronic coupling effects in the squaraine molecule. For this, we calculate Huang-Rhys factors $S_k$ of the modes with frequencies $\omega_k$ for the simplified squaraine molecule in vacuum within the DUSHIN program[94] using the force constants for the ground and first excited state geometries of the molecule. Fig. 13a summarizes the results for the theoretical displacements of the S1 state. The strongest component is a 155 cm$^{-1}$ mode with a $\Delta =1.08$ ($S = 0.58$), corresponding to the symmetric stretching motion of the two arms relative to the squaric acid center. A butterfly mode at 13 cm$^{-1}$ is too low in frequency to be resolved experimentally. The next strongest mode at 573 cm$^{-1}$ ($\Delta =0.33$) is another symmetric stretching mode. The strongly displaced modes at 155 cm$^{-1}$ and 573 cm$^{-1}$ agree very well with the two dominant modes at 147 cm$^{-1}$ and 570 cm$^{-1}$ observed experimentally. In the higher-frequency range, the calculations show a multitude of modes around 1500 cm$^{-1}$ that are only weakly displaced by less than 0.14.

We now use these mode frequencies and their displacements for the simulation of a pump-probe map using a two-dimensional DHO model[48]. We numerically integrate the Liouville-von Neumann equation in Lindblad form to obtain the time-evolution of the density matrix of our system, non-perturbatively interacting with a pump and a delayed probe field.[20, 85, 95-97] Relaxation and dephasing processes are described using Lindblad formalism.[97, 98] The simulated pump-probe map, depicted in Fig. 13b, agrees convincingly with the experimental map in Fig. 12a. The simulated residuals show a striking agreement in amplitude and phase profile with those from the experiment. In particular, the amplitude of the oscillations of the residuals is a most sensitive marker for the mode displacement, which can be reliably retrieved provided that the time resolution of the experiment is sufficiently short. The excellent agreement between experiment and simulation strongly validates that the vibronic couplings in ProSQ-C16 can well be interpreted in terms of DHO models based on the Born-Oppenheimer approximation. These simulations highlight the high degree of quantitative predictive power of the quantum chemical calculations for the vibronic coupling in our squaraine molecule.

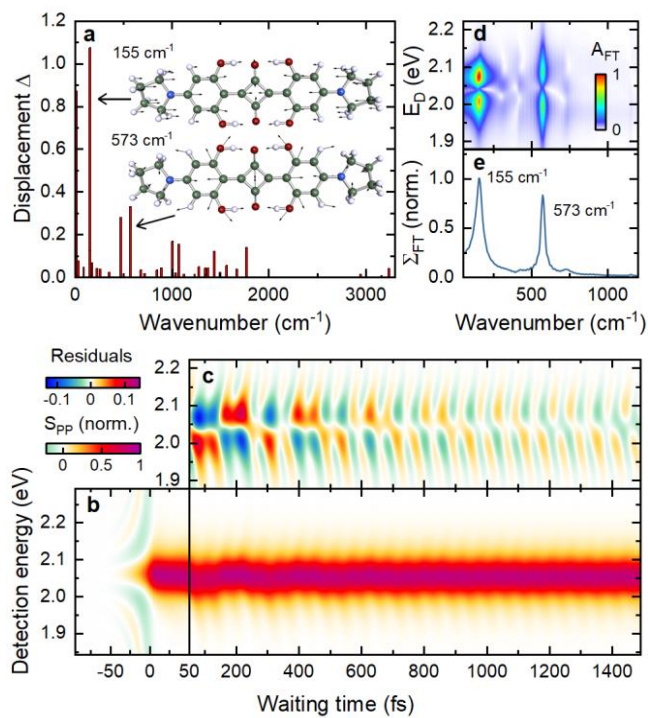

**Figure 13:** Theoretical modeling of vibronic couplings in a squaraine molecule. **a**: Dimensionless displacement $\Delta$ of the calculated vibrational modes for the squaraine molecule in vacuum. The insets show the spatial vibrational motion of the molecule for the two dominant modes at 155 cm$^{-1}$ and 573 cm$^{-1}$. The delocalization of the electronic wave function results in weak displacements of less than 0.14 for all high-frequency modes. **b**: Simulated pump-probe map based on the theoretical parameters for the two dominant vibrational modes and the electronic transition energy in chloroform. **c-e**: Residuals of the pump-probe map (**c**), its Fourier transform (**d**) and the vibrational spectrum (sum along the detection energy $E_D$, **e**) show good agreement with the experiment. These figures are reprinted with permission from Ref. [20], Copyright (c) 2016 APS Journals.

We argue that these comparatively small displacements of the high-frequency 1500 cm$^{-1}$ modes are an essential feature of the ESM in the parameter range that is relevant for our squaraine molecule. ESM predicts that the charge-delocalization in the D-A-D molecule results in a substantial reduction of vibronic coupling.[16] Within the ESM, the mixing of the symmetric superposition of diabatic CT states $|Z_+\rangle = (|Z_1\rangle + |Z_2\rangle)/\sqrt{2}$ with $|N\rangle$ creates a ρ-dependent quadrupolar moment of the adiabatic ground state $|g\rangle = \sqrt{1-\rho}|N\rangle + \sqrt{\rho}|Z_+\rangle$ due to the CT from the donor arms to the inner squaric acid acceptor.[16] In the case of $\rho = 0$, $|g\rangle$ has no CT character and its PES remains undisplaced along either the symmetric ($Q_+$) and antisymmetric ($Q_-$) coordinates. For a finite value of $\rho$, the symmetric CT shifts the equilibrium position of the molecular geometry and induces a displacement along the symmetric mode of $\Delta_{g,+} = \rho\Delta_{dia}/\sqrt{2}$. The optically excited $|c\rangle$ state has a $\rho$-independent displacement along $Q_+$ of $\Delta_{c,+} = \Delta_{dia}/\sqrt{2}$. Both states remain undisplaced along $Q_-$. Thus, the $|g\rangle$ and $|c\rangle$ PES are displaced by $\Delta_{gc,+} = (1-\rho)\Delta_{dia}/\sqrt{2}$ relative to each other. For our molecule with $\rho = 0.37$, this corresponds to a reduction in displacement in the quadrupolar molecule by a factor of 2.2 with respect to $\Delta_{dia}$, the displacement of the constituent polar dye. ESM thus predicts a reduction in Huang-Rhys factor by a factor of 5 due to the charge delocalization in the squaraine molecule.

In summary, our experimental results show that the low-energy optical excitations of the molecule are governed by a single one-photon allowed S1 state. A higher-lying two-photon allowed S2 state is energetically well separated from S1 by more than one eV. Time-resolved vibrational spectroscopy reveals an exceptionally weak vibronic coupling of the S1 state to the high-frequency stretching modes of the molecular backbone with Huang-Rhys factors of less than 0.03. The vibronic couplings in this quadrupolar molecule are dominated by a small set of lower frequency modes in the 100 to 600 cm$^{-1}$ range. Quantum chemical calculations firmly support these conclusions and provide quantitative striking agreement with experiment. Thus, quadrupolar squaraine molecules show interesting and unconventional optical properties. Their low-energy spectra are governed by a single excited electronic state with a large optical dipole moment and much reduced vibronic couplings to high-frequency carbon backbone modes. From a physics perspective, this makes such squaraine molecules highly interesting for applications in strong and ultrastrong coupling, in particular when further suppressing vibronic couplings by aggregation of squaraines in molecular thin films, as discussed in the next Section.

## 5. J-aggregated thin films of squaraine molecules and their couplings to plasmons

Based on the analysis of the electronic properties of squaraine molecules in solution we now turn our attention to J-aggregated squaraine thin films. Films with a thickness of 10 nm are formed by spin coating a solution of squaraine molecules[15] onto a glass substrate that was partially coated with a 200-nm thick gold film (Fig. 14).[20] To form J-aggregated molecular chains, the sample was further annealed on a hotplate.[15]

Such molecular aggregates are highly interesting supramolecular assemblies of non-chemically bound chromophores. Strong intermolecular Coulomb interaction between the chromophores delocalizes the excitonic wavefunction across several units.[99-101] This exciton delocalization leads to larger oscillator strengths of the aggregates ($|\mu|^2 = N_c|\mu_0|^2$) which scale with the number of coherently coupled chromophore units $N_c$,[102] compared to that of a single chromophore $|\mu_0|^2$. The same enhancement is seen for the radiative decay rate, known as superradiance.[103] Since the near-field dipolar coupling depends on both the energetic and geometric ordering of the individual chromophores, the collective aggregate excitations are highly sensitive to the molecular arrangement on the nanoscale.[104] Exciton delocalization, thus, allows for enhancing light absorption and tailoring excitation transport, making molecular aggregates of high fundamental and technological interest, *e.g.*, in natural light harvesting[105] and solar energy conversion.[99, 100]

Such aggregated molecular excitons inherently exhibit strong optical nonlinearities due to coherent wavefunction delocalization and exciton-exciton interaction[106-108] that can even facilitate nonlinear optical switching at the single-photon level.[109, 110] Importantly, molecular aggregates show often only one dominant electronic (excitonic) band at room temperature. Weak vibronic coupling to intramolecular vibrations,[111] resulting from the delocalization of the exci-ton wavefunction, also narrows this absorption band. These properties make molecular aggregates interesting for exploring strong coupling to vacuum fields[32, 112] or for enabling the study of quantum optical phenomena at room temperature[110].

The collective optical and electronic properties of molecular aggregates depend strongly on the exciton coherence length ($N_c$). Typically, organic thin films inherently suffer from static and dynamic disorder that limits the coherence length of molecular excitons. Such disorder is induced by spatial variations of individual chromophore resonance energies and by the coupling to vibrational degrees of freedom, respectively.[101, 113, 114] In the optical spectra, disorder manifests as an inhomogeneous broadening of the macroscopic optical response resulting from ensemble averaging over domains with different size and local structure. Therefore, a central research task is to increase the exciton delocalization length. The most obvious strategy is to improve the molecular packing of the aggregates by chemical manipulation.[13, 14, 115, 116] Alternatively, physical means, specifically the coupling of molecular excitons to localized electromagnetic field fluctuations, *e.g.*, in microcavities or localized plasmons at the surface of metallic nanoparticles, may enhance exciton coherence lengths and energy transport properties in molecular aggregates.[117-122] Here, our motivation is to investigate these physical means for exciton delocalization using 2DES.[64, 123] We expect that 2DES is a particularly powerful tool for such studies since it can naturally delineate homogeneous and inhomogeneous line broadening effects.[124-132]

## 5.1. Linear optical properties of J-aggregated squaraine thin films

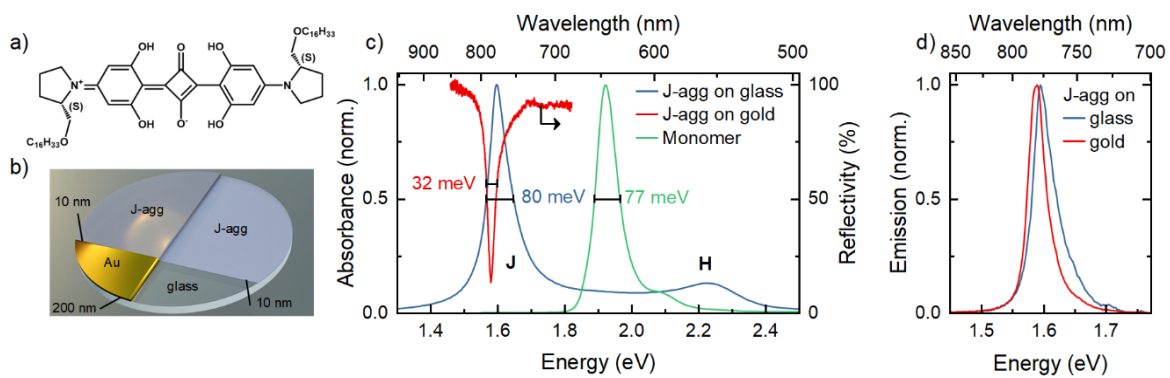

**Figure 14:** Sample and linear optical properties. (a) Structure of the squaraine molecule ProSQ-C16. (b) Schematic of the sample. An aggregated thin film of squaraines with a thickness of ~10 nm was prepared on a glass substrate, partially coated with a 200-nm thick gold film. (c) Absorption spectra of ProSQ-C16 monomers in chloroform (green) and of the aggregate on the bare glass substrate (blue). J and H bands of the aggregate can be observed in the latter. The reflection spectrum of the aggregate on the gold film (red) shows a reduced linewidth compared to the absorption on glass. (d) Fluorescence spectra of the aggregate on glass and gold. These figures are reprinted with permission from Ref. [20], Copyright (c) 2016 APS Journals.

In the experiments reported in this Section, we take squaraines as prototypical quadrupolar molecules for exploring plasmon-enhanced exciton delocalization in molecular aggregates.[20] Figure 14c compares the absorption spectrum of the aggregated film that has been deposited on the glass substrate (blue) with the reflection spectrum on of the aggregated film on a 200-nm thick gold film (red). Compared to the monomer absorption (green), both spectra show a red-shifted J-aggregate peak at ~1.6 eV.[15] On glass, a weaker, blue-shifted H-aggregate peak at ~2.22 eV is also visible.[101] We find that the J-aggregate band on gold (Jagg/Au) is only slightly red-shifted by about 16 meV compared to that on glass (Jagg/glass). This finding that the energies of the two J-aggregate bands are close suggests similar nearest neighbour coupling in both sample regions.[101] This implies that the assembly and formation of aggregates, mainly governed by intermolecular π-π or van der Waals interactions, depend weakly on the choice of the substrate. Importantly, a clear difference of the linewidth can be seen in Fig. 14c, with a full width at half maximum (FWHM) of ~32 meV for Jagg/Au compared to 80 meV for Jagg/glass. Since the two regions of the sample have comparable film thickness and peak positions of the J band, the difference of the linewidths suggests that a coupling to the substrate may affect the optical properties of the aggregates. Fluorescence spectra from both regions upon excitation at 633 nm are shown in Fig. 14d. The linewidth of the fluorescence spectrum of Jagg/glass is reduced by half. This indicates that the excitation at higher energies has relaxed towards the states at the bottom of the J band via intra- or inter-segment relaxation channels, from which radiative decay occurs.

## 5.2. 2DES of J-aggregated squaraine thin films

The distinct spectral linewidths of the linear absorption (reflection) spectra of Jagg/glass and Jagg/Au suggest a different exciton coherence length, despite the similar intermolecular electronic coupling strength in both regions. The linear spectra alone, however, cannot distinguish between homogeneous and inhomogeneous broadening. We therefore use 2DES to study the degree of disorder of the squaraine aggregates in the two different substrate regions. The 2DES measurements are performed with broad-band 12-fs pulses, centered at 780 nm, at a repetition rate of 175 kHz. We use a collinear, phase-locked pair of excitation pulses, delayed by the coherence time $\tau$, with a spectrum covering the entire J-aggregate resonance. Absorptive 2DES maps $A_{2D}(E_{det}, T, E_{ex})$ are recorded as a function of excitation ($E_{ex}$) and detection ($E_{det}$) energy.[85, 130] Fig. 15 and b show experimental 2DES maps at waiting time $T = 0$ fs for Jagg/glass and Jagg/Au, respectively. The general line shapes of both maps show clear dispersive peak features along the detection axis, in accord with earlier results in the literature.[124-129, 131] As demonstrated in seminal work by Wiersma's group[107], the peaks arise from the interplay between one-exciton and two-exciton transitions in the aggregate. Positive peaks (red) reflect ground state bleaching (GSB) and stimulated emission (SE) of the ground state $|0\rangle$ to one-exciton state $|X\rangle$ (0-X) transitions. Blue-shifted, negative peaks (blue) correspond to excited state absorption (ESA) from $|X\rangle$ to higher-lying two-exciton states $|XX\rangle$ (X-XX) in the aggregates.

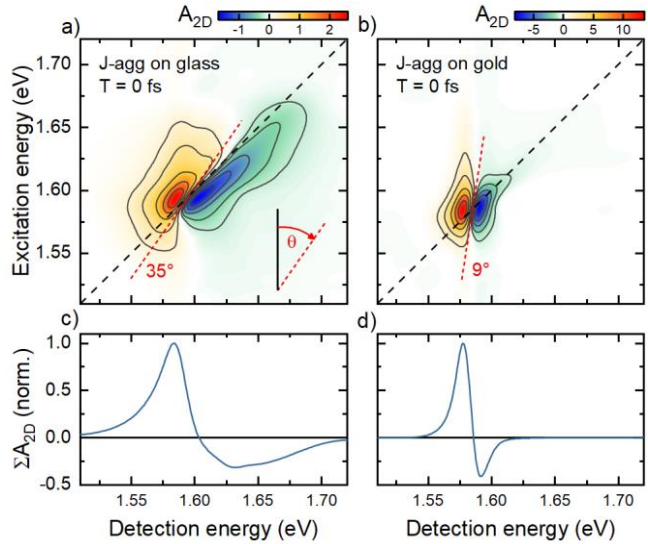

**Figure 15:** 2DES of the squaraine aggregates. (a,b) Experimental 2DES maps of the aggregate on the glass (a) and gold (b) regions of the substrate at waiting time $T = 0$ fs. Positive (red) signals correspond to ground state bleaching and stimulated emission whereas negative (blue) signals reflect excited state absorption. The spectra on glass show a large tilt angle (**θ**) of the zero-crossing line (red dashed) with respect to the excitation axis and elongated line shapes along the diagonal direction, indicating a finite amount of inhomogeneous broadening. This inhomogeneous broadening, and the tilt angle, is reduced for the aggregate on the gold film. (c,d) Integrated 2DES spectra along the excitation energy axis for (c) glass and (d) gold data highlighting the different spectral line shapes. These figures are reprinted with permission from Ref. [20], Copyright (c) 2016 APS Journals.

Even though both 2DES maps show these common peak features, their line shapes differ dramatically. For Jagg/glass, both the positive and negative peaks are elongated along the diagonal, which is a characteristic of inhomogeneous broadening in 2DES spectra.[133] For Jagg/Au spectra, however, the peaks are much narrower and mostly vertically distributed. The sizeable difference in peak shape between the two sample regions is further evidenced in the integrated 2DES spectra along the excitation axis shown in Fig. 15c and d. The ratio between inhomogeneous and homogeneous broadening in 2DES maps is usually characterized by the difference in linewidth between the diagonal and anti-diagonal spectra.[133] Such an analysis is complicated in our case because of the superposition of positive and negative signals and the existence of peak tails extending to higher excitation energies. Therefore, we take the angle $\theta$ between the zero-crossing line, separating positive and negative peaks (Fig. 15, red dashed line), and the excitation axis as a measure of the inhomogeneity. This angle reduces from

$\theta_{\text{glass}} \sim 35°$ for Jagg/glass to $\theta_{\text{Au}} \sim 9°$ for Jagg/Au. Specifically, for a single, homogeneously broadened exciton transition, the absorptive 2DES map would show Lorentzian, star-shaped peaks for both GSB/SE and ESA with linewidths determined by the exciton electronic dephasing rate and the angle $\theta$ would approach zero.[125] The 2DES map of Jagg/Au is, therefore, much closer to a homogenously broadened pattern, whereas the Jagg/glass map suggests a larger sample inhomogeneity, pointing to an increased disorder of the squaraine aggregates on glass.

In a simplified model, molecular aggregates can be considered as a one-dimensional, disordered chain of $N$ monomers (two-level molecules) and described by the Frenkel exciton Hamiltonian[101]

$$H_0 = \sum_{n=1}^{N} E_n |n\rangle\langle n| + \sum_{n,m=1}^{N} J_{nm} |n\rangle\langle m|, \tag{14}$$

where $E_n$ is the electronic excitation energy of the $n$-th monomer in the chain and $J_{nm}$ is the electronic coupling strength between sites $n$ and $m$. Here, we consider only the dominant nearest-neighbour coupling[134] and further assume identical nearest-neighbour coupling strength throughout the chain for simplicity. In such a model, the interplay between electronic coupling and disorder results in a delocalization of excitonic states over several monomers that is characterized by the exciton coherence length ($N_c$). $N_c$ measures the extent of the lowest-lying excitons, with s-like wavefunction character. In J-aggregates, spatial variations in the environment of each monomer and coupling to vibrations usually limit $N_c$ to values of the order of 10 at room temperature.[115, 124] In the Frenkel exciton model, this is included through random variations of the site energy, usually reasonably well described by a Gaussian distribution with standard deviation $\sigma$ and centred at the monomer absorption energy $E_0$ (Fig. 14c). A characteristic effect of intermolecular electronic coupling is exchange narrowing in which the linewidth $W$ of the absorption spectrum of the aggregates scales as $W \propto \sigma/\sqrt{N_c}$. This implies that the energy distribution of the optically bright excitons narrows by roughly a factor of $\sqrt{N_c}$ via the delocalization of the exciton wavefunction over $N_c$ monomers with uncorrelated transition energy.[134, 135]

Such a Frenkel exciton model immediately explains the dispersive nature of the 2DES peaks. In this model, one-exciton states with wave number $k$ and energy $E_k$, $|k\rangle = \sum_{n=1}^{N} \varphi_{kn}|n\rangle$, arise from the coherent superposition of electronic excitations with amplitudes $\varphi_{kn}$ at different monomer sites, where $|n\rangle = b_n^+|0\rangle$ and $b_n^+$ is the creation operator for the excitation at site $n$.[136] Since the monomer excitations obey Pauli's exclusion principle, each monomer site can only be occupied once and all two-particle excitations $|mn\rangle = b_m^+ b_n^+|0\rangle$ involve two different monomer sites $m \neq n$. This leads to two-exciton states $|k_1 k_2\rangle = \sum_{m>n}^{N} (\varphi_{k_2 n} \varphi_{k_1 m} - \varphi_{k_2 m} \varphi_{k_1 n})|mn\rangle$ with energies $E_{k_1 k_2} = E_{k_1} + E_{k_2}$ and amplitudes $\varphi_{kn}$ which are given by the respective one-exciton states.[136, 137] A consequence of Pauli's exclusion principle is that the energy $E_{1,2}$ of the lowest-lying two-exciton state $|1,2\rangle$ is always larger than twice the energy $E_1$ of the lowest one-exciton state $|1\rangle$, $E_{1,2} = 2E_1 + \Delta E$. Hence, the lowest lying optically bright excitation $|k\rangle$ on each segment is lower in energy than the dominant transitions from $|k\rangle$ to the two-exciton states $|k_1 k_2\rangle$. This level repulsion is at the origin of the optical nonlinearity of molecular excitons in aggregates and results in the blue shift of the ESA peak relative to the one-exciton transition.

To provide a direct measurement of the two-exciton shift $\Delta E$, we make use of the phase-cycling capabilities of the TWINS interferometer and record 0Q and 2Q spectra of a squaraine thin film deposited under optimized aggregation conditions on gold at room temperature[57]. The data are depicted in Fig. 16. A cross Section along the excitation axis of an absorptive 1Q 2DES (blue line in Fig. 16g) peaks at 1.578 eV, the energy of the lowest-lying exciton transition $E_1$. In contrast, a cross Section along the excitation axis of the 2Q spectrum at a detection energy $E_1$ peaks at 3.159 eV (black line in Fig. 16g), giving a two-exciton shift of $\Delta E = 3$ meV. A careful analysis of the corresponding 1Q spectra suggests a homogeneous dephasing time of the one-exciton transition of $T_2 = 160$ fs ($\hbar\gamma = \frac{\hbar}{T_2} = 4$ meV),[57] implying that all reported 2DES maps are largely inhomogeneously broadened.

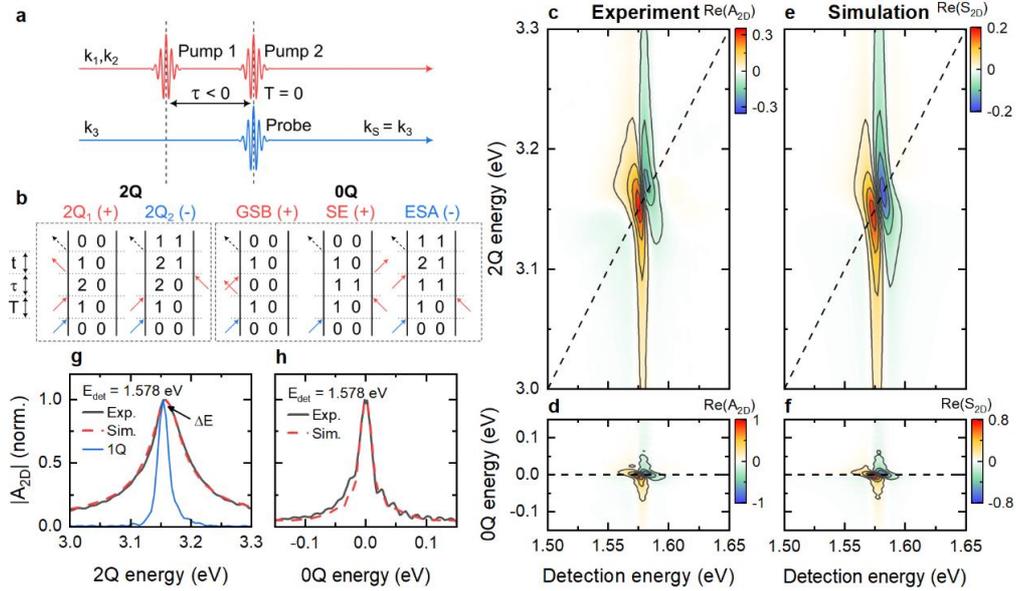

**Figure 16:** 2Q and 0Q spectra for a squaraine J-aggregate on a glass substrate. **a:** Experimental pulse scheme adapted for the acquisition of 0Q and 2Q spectra. By choosing $T = 0$ fs, the probe and second pump coincide. The first pump is now trailing since only negative coherence times $\tau < 0$ are scanned. **b:** Double-sided Feynman diagrams for the 2Q and 0Q spectra derived for the time-ordering in a. **c,d:** Real part of the experimental 2Q (c) and 0Q (d) spectra for the J-aggregate. **e,f:** Real part of the respective simulated spectra for a three-level system. **g:** Crosscut through the absolute-valued 2Q peak. A good agreement with the simulation is achieved when using a 2Q dephasing time $T_2^{2Q} = 30$ fs. A comparison to the absorptive one-quantum profile $|A_{2D}|$, shifted by $E_{det}$ along the excitation energy, allows to estimate the two-exciton shift $\Delta E$ of 3 meV. **h:** Crosscut through the absolute-valued 0Q peak. Using the simulation a time constant of $T_1 = 140$ fs is deduced. These figures are reprinted with permission from Ref. [20], Copyright (c) 2016 APS Journals.

### 5.3. Frenkel exciton simulations

To gain insight into the striking difference between the inhomogenous broadening of the aggregated film on gold and on glass, we fit the experimental 2DES maps by simulations based on the Hamiltonian in Eq. (14). While there are $N$ $|X\rangle$ and $N(N-1)/2$ $|XX\rangle$ states, only a few of them at the bottom of the one and two-exciton manifolds are optically bright.[106, 107, 136] Therefore, it is sufficient to only consider these low-energy transitions, particularly for calculating nonlinear spectra, to gain physical insights. For a homogeneous chain (inhomogeneous broadening $\sigma = 0$), the 0-X transitions with dipole moments $\mu_k = \mu_0 \sum_{n=1}^{N} \varphi_{kn}$ are dominated by the

$|k=1\rangle$ state. It carries 81% of the total oscillator strength since the transition dipoles $\mu_0$ of all monomers add up in phase. For the X-XX transitions at energy $E_{k_1k_2,k} = E_{k_1k_2} - E_k$, the dipole moment is $\mu_{k_1k_2,k} = \mu_0 \sum_{n_2>n_1}^{N}(\varphi_{kn_1} + \varphi_{kn_2})(\varphi_{k_1n_1}\varphi_{k_2n_2} - \varphi_{k_1n_2}\varphi_{k_2n_1})$ and, for $\sigma = 0$, the strongest contribution is from $|k=1\rangle$ to $|k_1=1, k_2=2\rangle$.[136]

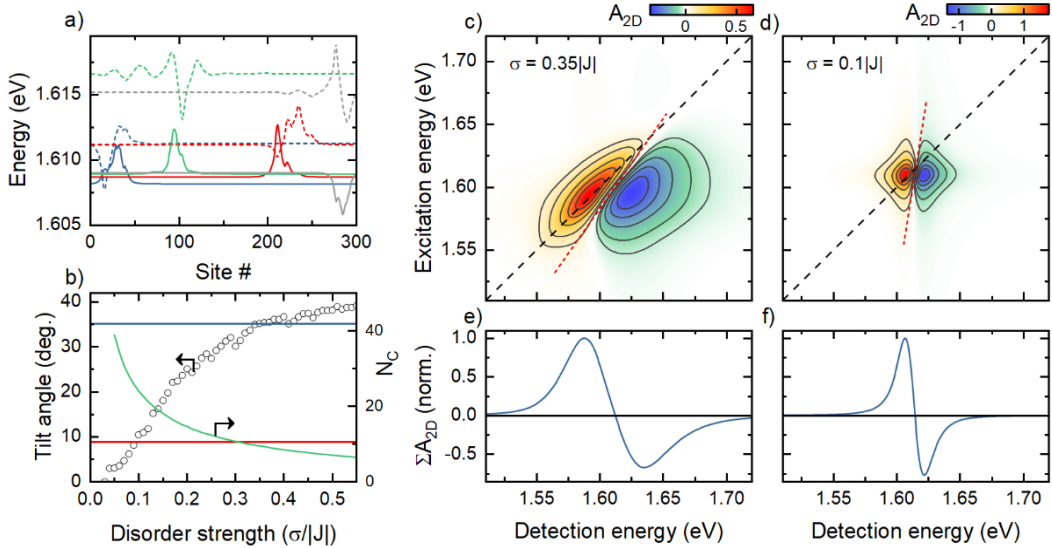

**Figure 17:** Simulation of the J-aggregate 2DES spectra using a Frenkel exciton model including disorder. (a) Wavefunctions of the four lowest-energy s-like (solid) and corresponding p-like states (dashed) for a specific realization of the disorder strength $\sigma = 0.1|J|$. The states with s- and p-like characters are localized on the same segment but separated in energy due to level repulsion. (b) Tilt angle of the zero-crossing line of the experimental 2DES map of the aggregate on glass (blue horizontal, $\theta_{\text{glass}} \sim 35°$) and gold (red horizontal, $\theta_{\text{Au}} \sim 9°$). Tilt angles obtained from simulated 2DES maps are plotted as black circles as a function of disorder strength. The corresponding number $N_c$ of coherently coupled molecules is shown as a green line. (c,d) Simulated 2DES maps for disorder strengths of $\sigma = 0.35|J|$ (c) and $0.1|J|$ (d) agree reasonably well with the experimental data. The corresponding integrated 2DES are shown in (e) and (f). These figures are reprinted with permission from Ref. [20], Copyright (c) 2016 APS Journals.

With increasing disorder, more and more states with node-less, s-like wavefunctions ($|s\rangle$-states) contribute to the optical spectra. These states are localized in different segments of the chain with an average localization length defined by $N_c$. This exciton localization leads to an inhomogeneous broadening of the 0-X peak as seen in 2DES. For ESA, only transitions from each localized exciton state $|s\rangle$ to a selected number of dominant two-exciton states is considered. Quantum level repulsion shifts their energy above that of $|s\rangle$, as seen from the representative wavefunctions of $|s\rangle$ (solid) and $|p\rangle$ (dashed) states from a specific disorder realization with $\sigma = 0.1|J|$ shown in Fig. 17. We follow the approaches in refs.[136, 138] to select the relevant states.

For computing 2DES maps, we numerically diagonalize the Hamiltonian in Eq. (14) and obtain $E_k$ and $\mu_k$ for each one-exciton state for a chain with a length of $N = 300$.[136, 137] The corresponding energies $E_{l,k}$ and dipole moments $\mu_{l,k}$ of the selected X-XX transitions are then deduced. The absorptive 2DES maps are simulated as[44]

$$A_{2D}(E_{det}, E_{ex}) = \langle \sum_k \frac{|\mu_k|^2 \gamma_1}{(E_{ex} - E_k)^2 + \gamma_1^2} \left\{ \frac{2|\mu_k|^2 \gamma_1}{(E_{det} - E_k)^2 + \gamma_1^2} \right. \quad (15)$$

$$\left. - \sum_l \frac{|\mu_{l,k}|^2 \gamma_2}{\left(E_{det} - E_{l,k}\right)^2 + \gamma_2^2} \right\} \rangle$$

Here, $\gamma_1$ and $\gamma_2$ are the electronic dephasing rates of 0-X and X-XX transitions, respectively. The first term accounts for GSB and SE of the $k$-th one-exciton transition and the summation is restricted to the optically bright $|s\rangle$ states. The second term describes ESA by the sum over the selected X-XX transitions. To model spectra with finite disorder, we calculate 2DES maps by averaging over 1000 realizations for each σ (denoted by $\langle \ \rangle$). In the simulations, we assume $\gamma_1 = \gamma_2 = 9$ meV, In this model, $\sigma$ remains as the only free fitting parameter, and thus, we compare 2DES maps with different $\sigma$ to the experimental results. For this, we take a fixed nearest neighbour-coupling $J = -154$ meV, determined from the energy difference between the J and H bands ($\sim 4|J|$).[101, 134] This implies that we only consider the optically dominant J-aggregate band in the 2DES simulations and also neglect spatial fluctuations in $J$ of the sample. The simulated values of the tilt angle $\theta_{sim}(\sigma)$ are plotted as black circles in Fig. 17b. Importantly, $\theta_{sim}$ monotonically increases with $\sigma$, which allows us to estimate $\sigma$ for the studied sample by comparing simulation and experiment. The latter are plotted in Fig. 17b as horizontal blue and red lines for Jagg/glass and Jagg/Au, respectively. From this comparison we determine $\sigma_{glass} \sim 0.35|J|$ and $\sigma_{Au} \sim 0.1|J|$. Simulated 2DES maps for these two disorder strengths are shown in Fig. 17c and d, reasonably well reproducing the experimental data in Fig. 17a and b, respectively. The integrated 2DES spectra in Fig. 17e and f also qualitatively fit the experimental results. Finite differences between experimental and simulated 2DES line shapes can in principle result both from the simplified disorder model, including only site disorder whereas $J$ is fixed, or from the restriction of Eq. (15) to the low-lying $|s\rangle$ states. We find that inclusion of higher lying states does not change the 2DES spectra significantly, leaving the simplified disorder model as the main cause for those differences.

The fitting results from this Frenkel exciton model provide an intuitive understanding of the reduced inhomogeneous broadening. It suggests that the disorder strength is reduced by more than three times when preparing the squaraine aggregates on a gold film instead of glass. The decrease of $\sigma$ leads to a larger extent of the excitonic wavefunction, and an increase in the number of coherently coupled monomers, $N_c$. Different methods have been used to estimate $N_c$[111, 114, 125, 135, 137, 139]. We determine $N_c$ from the inverse participation ratio $\mathrm{IPR}(E = E_k) = \sum_{n=1}^{N} \varphi_{kn}^4$, which characterizes the extent of a wavefunction at energy $E_k$.[137, 140] For a one-dimensional exciton model, the delocalization length is taken as $N_c(E) = \langle \frac{3}{2 \times \mathrm{IPR}(E)} - 1 \rangle$.[113, 137, 140] In the presence of disorder, $N_c(E)$ is energy dependent, being smallest for $|s\rangle$ states at the bottom of the J-band and increasing for those states closer to the centre of the one-exciton band. By averaging $N_c$ over the s-like states, we obtain a localization length for those excitons that contribute predominantly to the 2DES spectra. From the determined $\sigma_{\mathrm{Au}}$ and $\sigma_{\mathrm{glass}}$ we estimate $N_{c,\mathrm{Au}} \sim 24$, which is about 2.4 times larger than that on glass $N_{c,\mathrm{glass}} \sim 10$. The size of $N_c$ on glass is typical for molecular aggregates at room temperature.[124] An alternative way to calculate $N_c$ is to use the energy spacing of the $|s\rangle$ and $|p\rangle$ states, $\Delta E = \langle E_{|p\rangle} - E_{|s\rangle} \rangle$.[137] We obtain $\Delta E_{\mathrm{Au}}^{\mathrm{sim}} = 5.4$ meV and, correspondingly, $N_{c,\mathrm{Au}} = 28$ and $\Delta E_{\mathrm{glass}}^{\mathrm{sim}} = 29.2$ meV ($N_{c,\mathrm{glass}} = 12$), respectively. Both the absolute values and the ratio between $N_{c,\mathrm{Au}}$ and $N_{c,\mathrm{glass}}$ are comparable to the ones obtained using the inverse participation ratio, suggesting that the relevant eigenstates for localized excitons have been properly selected.

## 5.4. Plasmon-induced exciton delocalization

Experiment and simulation demonstrate that the exciton wavefunction of the molecular aggregates is spatially more extended when preparing the thin film on gold instead of glass. Since the preparation conditions and the resultant film thickness are the same, we expect a similar packing structure, and thus, a similar $\sigma$ of the aggregates on both gold and glass. Even though we cannot fully exclude the effect of different aggregate structures in the two sample regions, we argue that this arises from the coupling between J-aggregated excitons in the squaraine thin film and SPP excitations at the planar molecule-gold interface.[117-121, 141-143]

We use a phenomenological model to investigate the effect of SPPs on the optical properties of the aggregates. For this we treat intermolecular and molecule-SPP coupling on equal footing by adding a single, delocalized SPP mode $|P\rangle$ to Eq. (14) that couples to all monomers with the same exciton-SPP coupling strength $J_\mathrm{P}$.

$$H = H_0 + E_\mathrm{P}|P\rangle\langle P| + \sum_{n=1}^{N} J_\mathrm{P}(|n\rangle\langle P| + |P\rangle\langle n|) \qquad (16)$$

This means that each monomer, acting as an electric point dipole, can not only couple to the neighbouring molecules but can also excite the SPP mode with energy $E_\mathrm{P} = E_0$ at the molecule-gold interface via optical near-fields. The optical near-field of the propagating SPP along the interface can then further excite other molecules before it is damped. Effectively, the Hamiltonian in Eq. (16) describes the competition between disorder-induced exciton localization and exciton delocalization due to (i) nearest-neighbour interactions and (ii) coupling to the SPP mode. For simplicity, we assume identical orientations of all monomers and use a constant $J_\mathrm{P}$ value since the propagation length of the SPP mode, on the order of a few micrometres, is much longer than the chain length of the aggregate. We take the same disorder strength $\sigma = 0.35|J|$ for both Jagg/glass and Jagg/Au, as discussed above. In the absence of SPP coupling ($J_\text{P} = 0$ eV), the optical spectra deduced from Eq. (16) reduce to those of the Frenkel exciton model since far-field light cannot directly excite the SPP modes. Since our model considers the coupling between neighboring monomers and monomers and plasmons on equal footing, it can describe the coupled system both in the weak and strong coupling limits. To distinguish the two limiting cases, the couplings induced by the Hamiltonian need to be compared to the strength of the system-bath interaction. In our case, we have assumed that the system-bath coupling is weak and can be described by the dephasing rate $\gamma$.

Fig. 18a shows the wavefunctions of lowest lying s-like states, $|s_i\rangle$, in the absence of exciton-SPP coupling, indicating their localization in different segments of the chain with an extent of about 10 monomers. A finite coupling strength of $J_\text{P} = 0.028 J = -4.3$ meV results in wavefunctions displayed in Fig. 18b. Now, a few of the $|s_i\rangle$ states, that are close in energy but spatially separated, couple to each other and form superposition states $|j\rangle = \sum_i c_{ji}|s_i\rangle$, with $c_{ji}$ being the amplitude of the original state $|s_i\rangle$ in the wavefunction of $|j\rangle$. Here, we assume that the aggregate structures are one-dimensional chains and the coupling induced by SPP is anisotropic, predominantly along the chain direction. Importantly, the characteristics of the wavefunctions of these $|j\rangle$ states differ from the typical localized exciton states since they are delocalized over several, spatially well-separated segments of the aggregate. Thus, exciton-SPP coupling effectively bypasses disorder-induced Anderson localization and delocalizes the excitonic wavefunction over roughly $F_\text{s} \sim 2$ segments with considerable amplitudes $c_{ji}$, as schematically depicted in the insets of Fig. 18a and b. Quantitatively, we obtain $N_\text{c} \sim 20$ in the presence of exciton-SPP coupling using the inverse participation ratio of the wavefunctions of these coupled $|j\rangle$ states. This value is consistent with the $N_\text{c,Au} \sim 24$ deduced from the disordered exciton model.

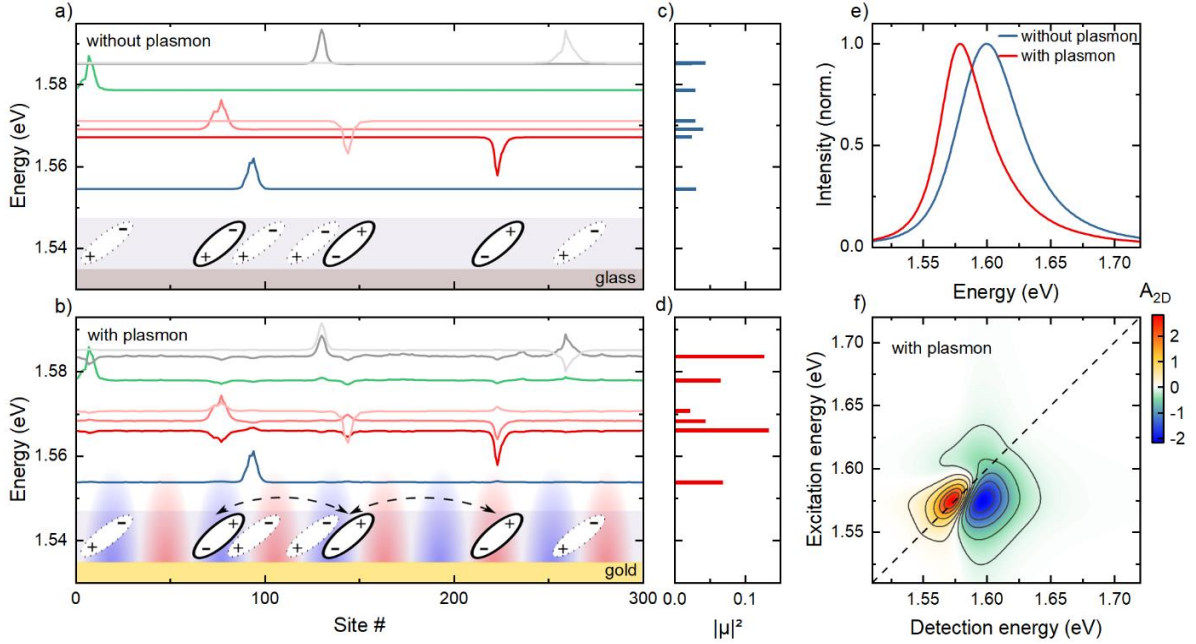

**Figure 18:** Plasmon-enhanced delocalization of exciton wavefunctions in J-aggregates. (a,b) Wavefunctions of the lowest-energy s-like states for a specific realization of the disorder ($\sigma = 0.35|J|$) without (a) and with (b) coupling to surface plasmon polaritons. Via dipolar exciton-plasmon coupling, several spatially separated but energetically close s-like states form superposition states with much more delocalized wavefunctions, as schematically shown in the insets of (a) and (b). (c,d) Oscillator strength $f = |\mu|^2$ of the states that are shown in (a) and (b). This coupling leads to the formation of a few bright, delocalized s-like states with enhanced oscillator strength, and to darker p-like states. (e) Simulated linear absorption spectra for $\sigma = 0.35|J|$ without (blue) and with exciton-plasmon coupling (red). (f) Simulated 2DES map for $\sigma = 0.35|J|$ in the presence of exciton-plasmon coupling, showing reduced inhomogeneity. The exciton-plasmon coupling strength is set to $J_P = 0.028J$. These figures are reprinted with permission from Ref. [20], Copyright (c) 2016 APS Journals.

The simulations show that the most significant effect mediated by the SPP mode is to induce a coupling between excitons with s-like character ($|s_i\rangle$ states) that are spatially localized in different segments of the aggregate. For our 2DES spectra, the coupled states with in-phase interference of $|s_i\rangle$ state wavefunctions are most relevant. The resultant enhanced oscillator strengths of those superradiant states profoundly affect the nonlinear 2DES spectra in Fig. 15. We see that the peak amplitudes of both GSB/SE and ESA of Jagg/Au are about 5 times stronger than that of Jagg/glass. This implies that not only 0-X but also X-XX transitions are enhanced. Theoretically, the nonlinear signal scales with $|\mu|^4$ and would be enhanced by a factor of $F_S^2 \sim 4$, which agrees well with the experiment.

Our analysis, therefore, suggests that the observed line width narrowing, reduction of inhomogeneous broadening and red-shift of the 2DES spectra, as well as the enhancement of the nonlinear signal strength for the aggregate on gold can indeed be well explained by including exciton-plasmon coupling effects. We note that these changes cannot be accounted for more simply by a slight variation the strength of the the nearest neighbor coupling $J$. Such modifications in resonance have a little effect on the line width.

In summary, the experiments that are discussed in this Section demonstrate an easy yet effective way to modify the electronic and optical properties of molecular aggregates via plasmon-enhanced exciton delocalization. 2DES measurements of squaraine-type aggregates reveal a significant reduction of inhomogeneous line broadening, providing direct spectroscopic evidence for the reduction of effective disorder strength when depositing the aggregates on gold. The experiments point to an enhancement of the exciton coherence length by roughly 2.4 times when switching from a glass substrate to the gold film. Frenkel exciton model calculations reproduce the experimental results reasonably well when including exciton-SPP coupling effects. The simulation suggests that the delocalized SPP mode at the molecule-gold interface leads to the coupling of a few energetically close but spatially well separated, quasi-localized excitons, forming delocalized, coupled super-radiant states. Importantly, the Frenkel-exciton simulations treat the J-aggregated excitons as purely electronic excitations, neglecting vibronic couplings to the vibrational modes of the molecular constituents. The recorded 2DES maps therefore provide convincing evidence that exciton delocalization in aggregates with strong nearest neighbor near-field dipole coupling further suppresses the already weak vibronic couplings in the quadrupolar squaraine molecules. This makes squaraine-based J-aggregates interesting model systems for delocalized excitonic quasi-three-level systems with large oscillator strengths and comparatively long excitonic dephasing times, exceeding 100 fs at room temper-ature. Residual inhomogeneous broadening, resulting from unavoidable disorder in the aggregated thin film, is readily suppressed by sufficiently strong near-field dipole couplings to delocalized cavity modes, as exemplified here for the dipole coupling to propagating surface plasmon polariton modes of a metallic film. This makes such aggregates of quadrupolar squaraine dyes very attractive as active quantum emitters for (ultra-)strong coupling to vacuum fields and Bose-Einstein Condensation applications, *e.g.*, for ultrafast and efficient switching of light by light on the nanoscale. The delocalization of exciton wavefunctions via coupling to SPPs, creating inter-segment, long-range coherent excitons, will also help to improve the coherent exciton energy transport properties of organic thin film nanostructures for optoelectronic applications.

## 6. Strong exciton-plasmon coupling dynamics in squaraine/gold nanoslit hybrid nanostructures

In this Section, we make use of the knowledge gained about the electronic and optical properties of squaraine molecules in solution and in J-aggregated thin films to explore the strong coupling of J-aggregated squaraine excitons to surface plasmon polariton (SPP) excitations of a gold nanoslit grating. For this, we set up angle-resolved 2DES experiments with the major aim to probe and unambiguously identify coherent Rabi oscillations between excitons and SPPs in the form of temporally oscillating cross peaks in the 2DES data. We indeed observed the anticipated ultrafast cross peak oscillations yet learned, much to our surprise, that they do not only arise from a coherent energy transfer between excitons and SPPs but reflect a coherent long-range energy transfer between excitons in different regions of the nanostructure, mediated by the plasmonic field. The Section summarizes the experimental findings and discusses the underlying physics.

## 6.1 Experimental result

To probe strong couplings between excitons and plasmons, we performed angle-resolved 2DES on a hybrid plasmonic cavity, a gold nanoslit array covered with a J-aggregated thin film (Fig. 19). This molecular aggregate is based on the squaraine monomers discussed in Section 4 . As shown in Section 5, their electronic properties are well described within ESM[19] and only the lowest-lying electronically excited state is relevant for the present work. When deposited on gold, the squaraine molecules form well-ordered J-aggregated thin films[20, 144] in which dipolar coupling among neighboring molecules results in a delocalization of the optical excitation across ~ 20 - 30 monomers at room temperature and in the formation of strongly red-shifted and spectrally narrow superradiant exciton $|X\rangle$ resonances at around 1.59 eV[20] (Fig. 19b).

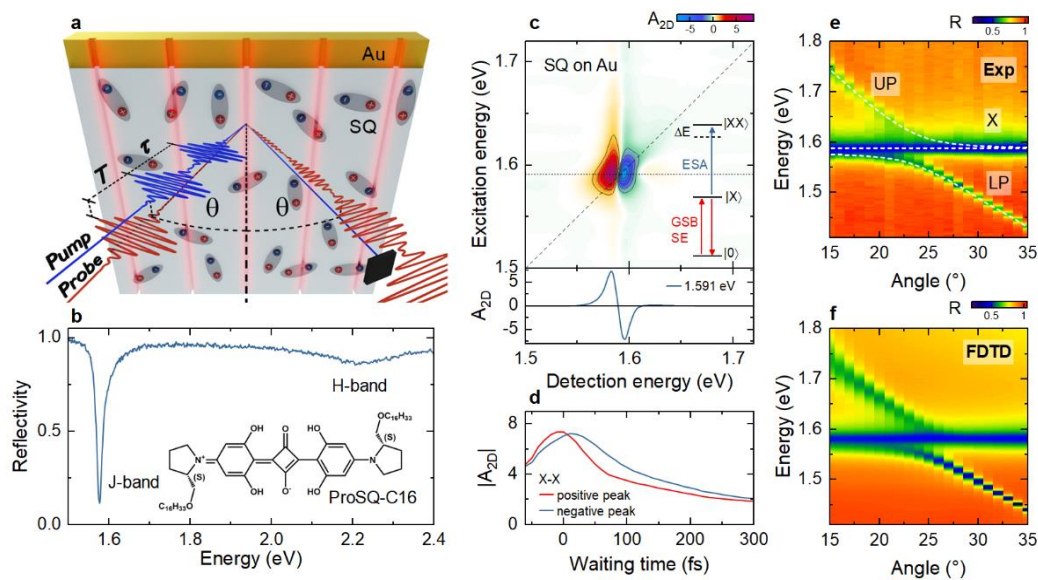

**Figure 39:** Strong coupling of a gold nanoslit array coated with a J-aggregated thin film. a) Experimental geometry. The nanoslit sample is illuminated with a phase-locked pair of pump pulses, separated by the coherence time $\tau$ at incidence angle $\theta$. The pump-induced change in sample reflectivity is monitored by a probe pulse with the same incidence angle and time-delayed by the waiting time T. b) Linear reflectivity of a 10-nm J-aggregated film of squaraine molecules (inset) on a gold surface. c) Experimental, reflective two-dimensional electronic (2DES) spectrum of the film on a flat gold surface at $T = 0$ fs. Bottom: Cross section along the detection energy for excitation at 1.591 eV showing the characteristic dispersive line shape of the J-aggregate exciton. Inset: One-quantum ($|X\rangle$) and two-quantum ($|XX\rangle$) excitations contributing to the 2DES exciton peak. d) Waiting time dynamics at the maximum (red) and minimum (blue) of the 2DES exciton peak, showing incoherent relaxation dynamics on a 100-fs time scale. e) Angle-resolved linear reflectivity spectra reveal the dispersion relations of upper (UP) and lower (LP) polaritons together with an angle-independent peak of "uncoupled" excitons (X). f)

Finite-difference time domain (FDTD) simulation of the angle-resolved reflectivity. These figures are reprinted with permission from Ref. [21], Copyright (c) 2016 APS Journals.

In 2DES, this results in a spectrally sharp and well isolated exciton peak with a dispersive line shape along the detection axis (Fig. 19c). The 2DES maps are recorded in reflection geometry. We label the peaks in these maps as ($Ex, Det$), where $Ex$ and $Det$ denote the excited and detected resonances, respectively. As discussed in Section 5, the dispersive line shape arises from a superposition of GSB and SE of the one-quantum $|X\rangle$ resonance which spectrally overlaps with ESA from $|X\rangle$ to two-exciton, $|XX\rangle$, states. The two-quantum $|XX\rangle$ resonances are blue-shifted ($\Delta E$) since Pauli-blocking in each monomer dictates that the lowest-lying delocalized exciton state in every aggregate can be populated only once[145]. Slight line broadenings result from higher-lying aggregated exciton states. Exciton relaxation within the disordered aggregates leads to a partial decay of the $|X\rangle$ peak on a 100-fs time scale (Fig. 19d).

We deposit such J-aggregated thin films, with a thickness of 10 nm, onto a plasmonic nanoslit array, milled into a 200-nm thick gold film. The nanoslits form a cavity that locally confines optical near fields and strongly enhances their coupling to the in-plane component of the exciton dipole moment. Width, height (45 nm) and period (530 nm) of the array are chosen to create sharp SPP resonances of the grating with an energy that can be tuned across the exciton resonance by varying the incidence angle $\theta$. Angle-resolved linear reflectivity spectra (Fig. 19e) show that the collective dipolar coupling between excitons and SPPs results in the formation of mixed upper (UP) and lower (LP) polariton branches. From the avoided crossing, we deduce a normal mode splitting of ~60 meV, twice the Rabi energy $\hbar\Omega_R$. The polariton branches are superimposed by an angle-independent exciton peak. It is commonly thought to arise from "uncoupled" excitons which are only weakly interacting with the SPP field, e.g., because they lie in regions outside the slits with much reduced local field enhancement[146, 147]. This interpretation of the linear spectra is well supported by finite difference time domain (FDTD) simulations of Maxwell's equations (Fig. 19f and Fig. 20).

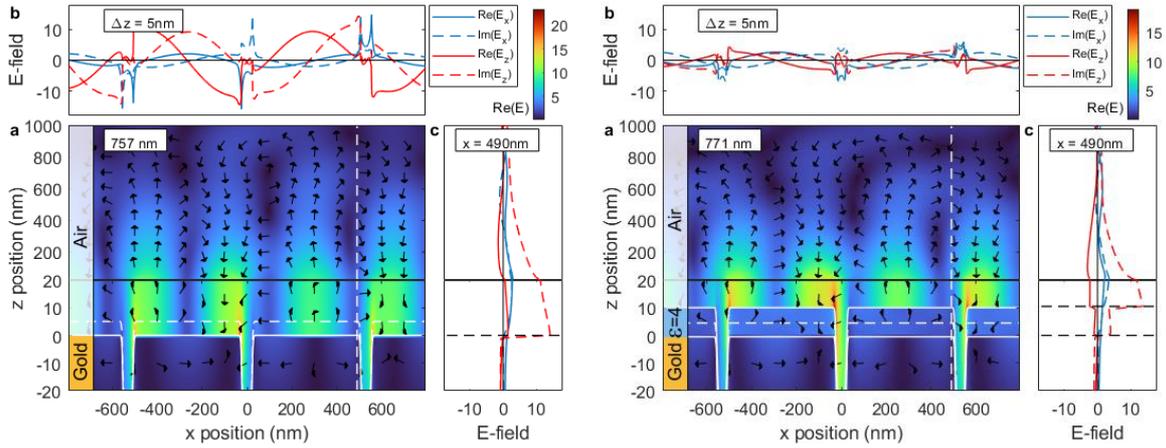

**Figure 20:** Electric field distribution across the interface between a structured gold film containing an array of nanoslits with 530 nm period and air. The simulations are performed for gratings without any coating (left) and with a 10-nm-thick dielectric coating with $\varepsilon_{Diel} = 4$ (right). The cross sections in b are taken at a constant height of $\Delta z = 5$ nm above the gold surface. The spatial field distributions are shown at the wavelength of the AM[-1] SPP resonance with a field enhancement of ~10. In addition, field enhancements are seen near the edges of the slits. Inside the dielectric coating, the z-component of the field is strongly reduced.[21] These figures are reprinted with permission from Ref. [21], Copyright (c) 2016 APS Journals.

Most of these resonances also appear in angle-dependent 2DES spectra recorded at $T = 0$ fs (Fig. 21).[21] Along their diagonal, we observe strong (LP,LP) and (X,X) peaks with dispersive line shapes along $E_{det}$. In contrast, the (UP,UP) peak is much weaker and appears only at angles below the crossing at $\theta_c = 23°$. In addition, we find pronounced cross peaks between LP and X, both below and above the diagonal. Their dispersive line shapes are best seen for $\theta = 27°$ in Fig. 21c. Cross peaks between UP and both, LP and X, are much weaker in amplitude and are only resolved for $\theta < \theta_c$. While (X,UP) and (UP,X) have dispersive line shapes, the other weak UP peaks appear absorptive in shape. Resonance energies and spectral line shapes deduced from 2DES are supported by angle-resolved pump-probe measurements.[21] The quite pronounced cross peaks between "uncoupled" excitons and polaritons are unexpected since the uncoupled excitons are thought to be spatially well separated and thus essentially uncorrelated with those excitons that are hybridized with the SPP field. Hence, it is not obvious that their excitation should result in a polariton nonlinearity.

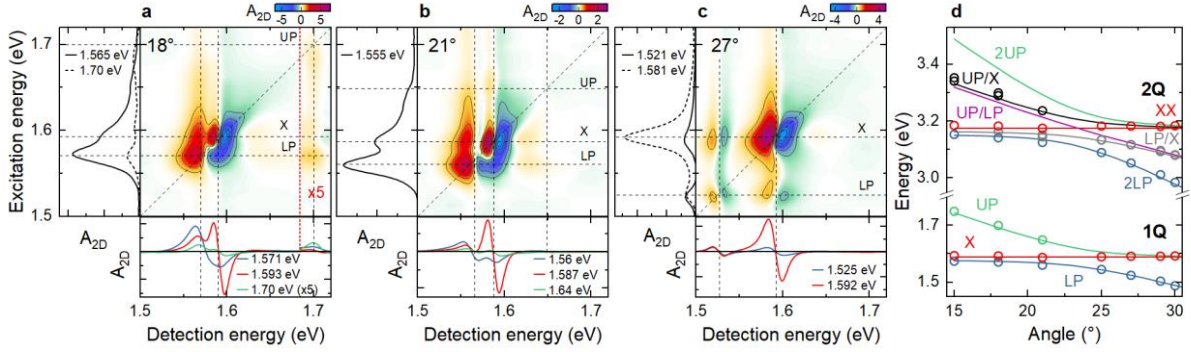

**Figure 21:** Experimental 2DES maps of the J-aggregate-nanoslit array for selected incidence angles at waiting time $T = 0$ fs. Insets: Cross sections at selected excitation and detection energies. a) 2DES map for $\theta = 18°$ displaying diagonal and cross peaks at the LP (1.571 eV) and UP (1.70 eV) energies. In addition to the dispersive "uncoupled" X peak (1.593 eV), cross peaks between both polaritons and the X transition are observed. The deduced resonance energies are marked as dashed lines. b) The same general features are also seen at $\theta = 21°$. The resonance energies of the polariton peaks are shifted according to their dispersion relation. c) For $\theta = 27°$, the detuning between LP and X is sufficiently large to resolve the dispersive line shape of all four diagonal and cross peaks of LP and X. At this angle, the intensities of the UP-related peaks are too weak to be seen. d) Dispersion relation of one-quantum (1Q) and two-quantum (2Q) excitations, as deduced from 2DES maps (open circles). The 1Q energies match well the linear dispersion relation (solid lines). 2Q excitations of lower polaritons (2LP) and "uncoupled" excitons (XX) are extracted together with mixed LP/X and UP/X peaks, while 2UP and UP/LP peaks are lacking. These figures are reprinted with permission from Ref. [21], Copyright (c) 2016 APS Journals.

The observation of dispersive line shapes for both diagonal and cross peaks now allows us to correlate one-quantum (1Q) resonances, characterized by a positive GSB and SE peak, and two-quantum (2Q) resonances with a negative ESA signal[20, 148]. We deduce 1Q energies from peak maxima along $E_{ex}$, while 2Q energies are taken as the zero crossing of a dispersive peak along $E_{det}$. The resulting energies are plotted in Fig. 21d as open circles, together with the 1Q dispersion (solid lines) deduced from angle-resolved reflectivity. In addition, the 2Q disper-sions are estimated by adding the energies of the contributing 1Q states without further corrections. Obviously, the 1Q energies obtained from 2DES match those deduced from linear spectroscopy, while the 2Q dispersions show several new features. Since the experiment probes the collective coupling of many excitons to a single plasmonic mode, we expect, from a commonly employed Tavis Cummings (TC) model, to observe three distinct 2Q states[148, 149]. The model predicts doubly excited 2LP and 2UP polaritons and a mixed UP/LP state, while all other states remain optically inactive ("dark")[148]. Indeed, these resonances have been seen in the 2Q dispersions measured for semiconductor microcavities[149] and a TC model has recently also been used to discuss organic microcavity polaritons[148]. As a result of the fermionic nonlinearity induced by the exciton part of the wavefunction, the energies of the 2Q states are slightly blue-shifted with respect to twice the 1Q transition. This blue shift is proportional to the two-exciton fraction of the 2Q wavefunction[148, 149]. Since the doubly excited X state is uncoupled from the plasmon branch, an angle-independent XX contribution is expected (Fig. 21).

In the experiments, the 2LP resonances are clearly resolved, while 2UP and UP/LP are apparently lacking. The mixed resonances (grey and black circles) follow the UP/X and LP/X dispersions with a distinct avoided crossing at $\theta_c$. The appearance of those resonances goes beyond the TC model and requires further discussion.

Before that, however, we inspect the waiting time dynamics of the 2DES spectra. This is exemplarily done in Fig. 22a for $\theta = 27°$. Indeed, we observe pronounced coherent oscillations of the amplitude on all diagonal and cross peaks in the 2DES map, except for the (X,X) peak. We will argue below that the oscillatory waiting time dynamics of the 2DES peaks provides crucial new information about coherent couplings in exciton-plasmon systems that has not been obtained from one-dimensional pump-probe spectroscopy, as previously reported by one of the present authors[30]. Interestingly, for $\theta = 27°$ (Fig. 22a), the oscillation period $T_X = 2\pi/(\omega_X - \omega_{LP})$ matches the splitting between the X and LP resonances. Coherent oscillations with the same period are also observed in pump-probe spectra (Fig. 22e).[21] The traditional understanding of strong X-SPP coupling would, instead, predict Rabi oscillations with a period $T_R = 2\pi/(\omega_{UP} - \omega_{LP})$ given by the normal mode splitting of the polaritons[30-32, 150]. Angle-dependent pump-probe transients, detected at the LP resonance (Fig. 22e), again reveal $T_X$ oscillations with a period that decreases monotonically with increasing angle. For $\theta \geq 21°$, also the 2DES maps show persistent amplitude oscillations, except for (X,X). These oscillations appear predominantly with a period given by the X-LP splitting, as can be seen by comparing the measured oscillation periods (blue squares in Fig. 22f) to those predicted by the linear dispersion (blue line). Only at one selected angle, close to the crossing, we find an oscillation at the anticipated UP-LP splitting (red circle in Fig. 22f).

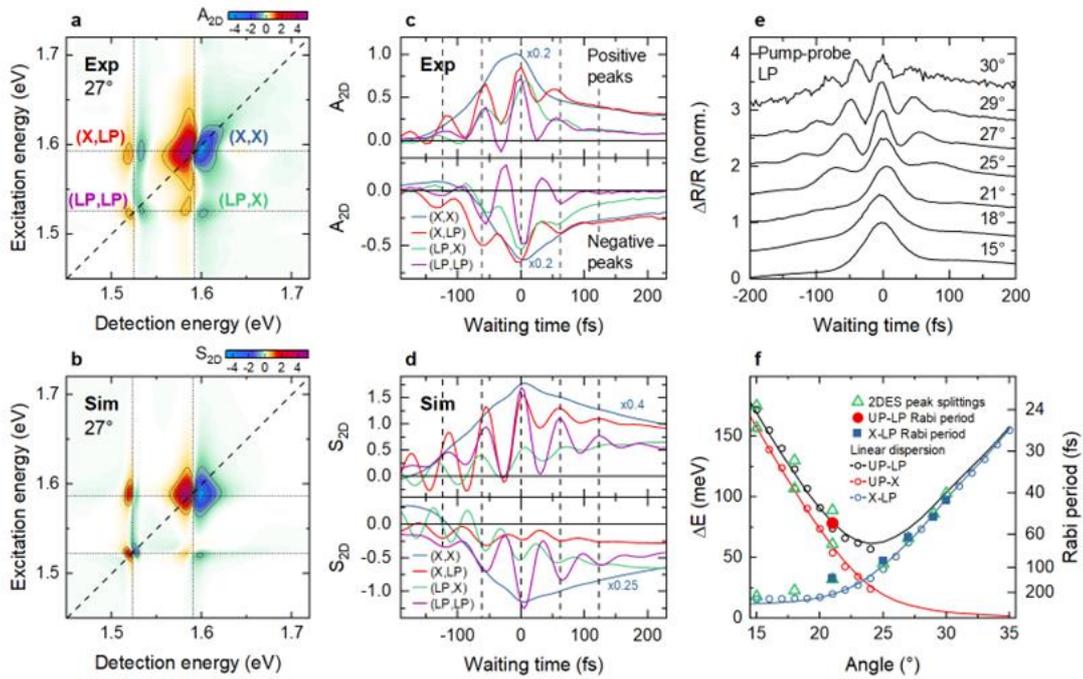

**Figure 22:** 2DES dynamics revealing polariton Rabi oscillations. a) Experimental 2DES map at $T = 0$ fs and $\theta = 27°$. b) Simulation of the 2DES map using the effective coupling Hamiltonian introduced in the text. As in the experiment, diagonal and cross peaks with dispersive line shape involving the LP and X transitions appear, while the UP-related peaks are too weak. c) The waiting time dynamics of the (LP,LP) diagonal and (LP,X) / (X,LP) cross peak reveal pronounced Rabi oscillations with a period matching the peak splitting while such oscillations are lacking at the (X,X) diagonal. The dynamics detected on the positive and negative sides of the dispersive peaks are displayed in the upper and lower panels, respectively. d) Simulations of the waiting time dynamics for all peaks shown in c). e) Angle-resolved pump-probe dynamics detected at the positive peak of the dispersive LP resonance. f) Rabi oscillations periods $2\pi\hbar/\Delta E$ (filled symbols) are extracted from the 2DES waiting time dynamics for different incidence angles. UP-LP oscillations (red circle) are only resolved at $\theta = 21°$, near the crossing angle. All other periods (blue squares) match the X-LP splitting. The 2DES peak splittings (green triangles) follow the energy differences $\Delta E$ between the corresponding UP, LP and X transitions, as deduced from the linear dispersion (open circles and lines). These figures are reprinted with permission from Ref. [21], Copyright (c) 2016 APS Journals.

## 6.2 Physical interpretation of the experimental results

To explain these experimental observations, we first introduce a phenomenological extension of the TC model that takes the spatial characteristics of our sample into account. For this, we consider two classes of spatially separated J-aggregated excitons. Excitons in the slit region, $X_S$, collectively interact with the plasmon field with a coupling strength $V_S$. In contrast, those excitons, $X_W$, that lie between the slits, on the flat gold film, interact with a smaller coupling strength $V_W$. Both plasmons and excitons are treated as bosonic oscillators[149]. A nonlinearity of the system arises by introducing a finite blue-shift $\Delta E$ of the two-exciton states of the strongly and weakly coupled excitons. Using this model, 2DES spectra are simulated by solving the Lindblad master equation for the density matrix of the coupled system. As can be seen in Figs. 21b,d, the model quantitatively accounts for our experimental observations. Specifically, the simulations show dispersive peaks with pronounced amplitude oscillations at the period $T_X$, given by the X-LP splitting, while Rabi oscillations at $T_R$ are much weaker in amplitude. As in the experiment, the oscillations are basically absent at (X,X). Reasonable agreement between experiment and simulation is achieved when choosing $V_S \simeq 3V_W$, with $\hbar\Omega_R = \sqrt{V_S^2 + V_W^2}$, and $\mu_P \simeq \mu_W \simeq 3\mu_S$.[21]

To rationalize the microscopic dynamical processes that give rise to these transient 2DES spectra we further invoke an elementary Frenkel exciton model[20, 145]. We consider a disordered chain of squaraine molecules, each treated as a fermionic two-level system. Neighboring molecules are dipole-coupled via their optical near fields and interact with the plasmonic mode that is delocalized along the chain. In agreement with FDTD simulations of our sample, we consider a spatially inhomogeneous plasmon field with a local field in the slit region that is ten times larger than in the region between the slits (Fig. 23a). In the absence of the plasmon field, the nearest-neighbor coupling results in the formation of few superradiant, moderately localized J-aggregated excitons, strongly red-shifted in energy and extending over ~25 molecules, together with a large number of dark excitons. Between the slits, the wavefunctions of these localized excitons ($X$) remain basically unchanged in the presence of the coupling to the plasmon mode – except for a minor admixture of excitons inside the slits. The plasmon contribution to their wave function is small. In contrast, the superradiant excitons inside the slits couple strongly to the plasmonic mode, resulting in a LP mode carrying substantial contributions from $X_S$ and $P$ and much weaker contributions from all $X_W$. For LP, all wavefunctions interfere constructively while for the $X$ states, the contributions from $X_S$ and $X_W$ interfere destructively. The resulting linear optical absorption (Fig. 23b) shows strong contributions from the energetically isolated LP state, while the X peak is inhomogeneously broadened. The UP absorption is much weaker since, in our sample, the dipole moment of $P$ ($\mu_P$) and of the sum of all excitons ($\mu_W$ and $\mu_S$) are of similar magnitude. Hence, their emission interferes destructively for the UP peak.

We now discuss the dynamics of the coupled X-SPP system. For this, we impulsively excite all optical resonances with a spatially homogeneous laser field and follow the spatiotemporal evolution of the excited state populations within the chain of squaraine monomers and in the plasmon mode. The plasmon mode shows the expected Rabi oscillations with $T_R$. Out-of-phase oscillations at $T_R$ are most pronounced for the slit excitons $X_S$. They are superimposed, however, with slower oscillations of the $X_S$ population with a period $T_X$. While these slower oscillations are completely absent in the plasmon dynamics, they reappear, phase-shifted by $\pi$, for those excitons, $X_W$, that are localized between the slits. These two distinct types of population oscillations are most clearly seen when spatially integrating over the localized exciton populations $X_S$ and $X_W$ (Fig. 23c). Now, the Rabi oscillations on $P$ are perfectly matched by out-of-phase oscillations of the total population of all excitons, $X_S + X_W$ (red line in Fig. 23c). The plasmon-mediated coherent population oscillations (CPO) between $X_S$ and $X_W$ are only seen in the individual exciton subsystems, while they are absent in the net exciton population.

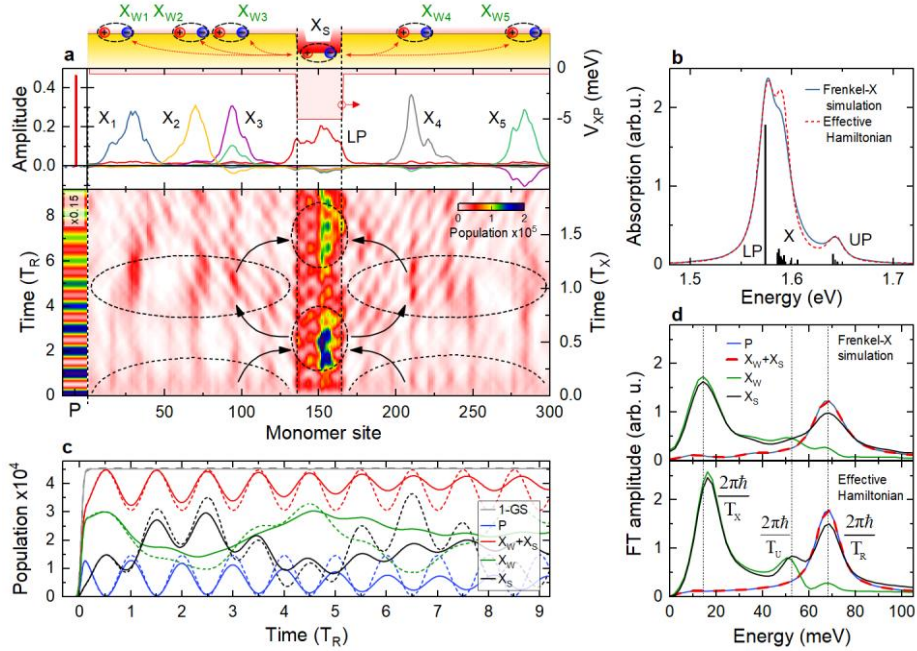

**Figure 23:** Frenkel exciton simulations of plasmon-driven coherent exciton population oscillations (CPOs). a) A chain of 300 squaraine monomers with disordered site energies and nearest-neighbor coupling forms localized J-aggregate excitons. The monomers are coupled to a delocalized plasmonic mode with a spatially inhomogeneous coupling strength $V_{XP}$ that decreases in amplitude from 5.0 meV for excitons ($X_S$) inside the narrow slit region to 0.5 meV for those ($X_W$) outside the slits. This leads to the absorption spectrum in (b), displaying a single, delocalized LP mode and several disordered UP transitions together with an "uncoupled" exciton peak X. Excitons and plasmon mode (P) are resonantly excited by a 5-fs pulse. After excitation, the population of P displays oscillations at a period $T_R = 2\pi/(\omega_{UP} - \omega_{LP})$, given by the UP-LP splitting. Out-of-phase UP-LP oscillations are seen for the slit excitons. These are superimposed by slower oscillations at $T_X = 2\pi/(\omega_X - \omega_{LP})$. These reflect CPOs from outside the slits into the slit region and back, as illustrated by the arrows (a). c) Population dynamics of the $X_S$, $X_W$ and $P$ states after impulsive excitation, displaying oscillations at $T_R$ and $T_X$. CPOs are seen on both, $X_S$ and $X_W$, but are fully absent for $P$ and $X_S + X_W$. Dynamics for the effective Hamiltonian are shown as dashed lines. d) Fourier transform of the populations displaying CPOs at $T_X$, Rabi oscillations at $T_R$, and weaker oscillations at $T_U = 2\pi/(\omega_{UP} - \omega_X)$. These figures are reprinted with permission from Ref. [21], Copyright (c) 2016 APS Journals.

These model calculations suggest that the dipolar coupling to the plasmon induces spatial oscillations in the exciton density, from outside the slits into the slit region and back. These oscillations appear at the period $T_X$, given by the energy splitting between X and LP. They clearly dominate the exciton dynamics in the region between the slits. Here, the effect of the "traditional" Rabi oscillations with period $T_R$ is weak due to the small local plasmon field amplitude. These conclusions are largely corroborated by Fourier transforms of the population dynamics (Fig. 23d). They emphasize the presence of fast Rabi oscillations with $T_R$ and absence of $T_X$ oscillations in the dynamics of the plasmons and of the total exciton population, $X_S + X_W$, (blue and red line), respectively. In contrast, the slower oscillations with $T_X$ between the two distinct classes of excitons become apparent when examining the individual exciton dynamics. These Frenkel exciton simulations form a convincing microscopic basis for the phenomenological extension of the TC model introduced above. Essentially, we can explain the suppression of "traditional" exciton-plasmon Rabi oscillations ($T_R$) and the emergence of CPOs with a longer period $T_X$ in the waiting time dynamics of the 2DES maps by considering two spatially distinct classes of localized J-aggregated excitons that are mutually coupled to a spatially structured plasmonic mode. This model accounts for the rich spectral and dynamic features in all angle-dependent 2DES maps. Most importantly, it provides a physically intuitive explanation how a spatially delocalized plasmon mode induces a coherent real-space energy transport between spatially separated exciton sites that persists during the coherence time of the strongly coupled system. This maps the complex real space dynamics of a nanostructured system of molecular excitons and plasmons onto an effective three-level system, in which two of the levels, $X_S$, the excitons inside the slits, and $X_W$, the excitons between the slits, are both coupled to a third state, the plasmon. The collective coupling strength of the excitons inside the slits is roughly three times larger than that for those outside the slits. This spatially structured coupling gives rise to CPOs between two of these states without affecting the dynamics of the third. Previously, such CPOs have been discussed in atomic and molecular three-level systems in the context of slow light generation[151] and light storage[152]. Here, we report CPOs in a prototypical all-solid 3-level-system at room temperature and demonstrate how they enable an efficient coherent transport of excitons over mesoscopic distances, from regions outside to inside the slits and back. Atomic and molecular 3-level and 4-level systems offer exciting resources for quantum state manipulation and information processing. The reduction in the speed of light by electromagnetically induced transparency[153], the coherent trapping of population in optically dark states by stimulated Raman adiabatic passage[154] or lasing without inversion[155] are among the manifestations of the control of optical information that can be achieved. We therefore anticipate that the demonstration of related coherent phenomena in all-solid-state systems will open up new avenues towards optical information processing in strongly coupled exciton plasmon systems. Our results show that strongly coupled exciton-plasmon systems offer exciting new prospects for manipulating coherent quantum transport by light. To leverage these opportunities, direct spatial and temporal visualization of the exciton transport dynamics will be an important next step.

## 7. Conclusions and Outlook

In conclusion, we believe that this chapter reports on a rather substantial advance in two-dimensional electronic spectroscopy that has been achieved in the laboratory of the authors during the past three years. Taking quadrupolar squaraine molecules and their thin film aggregates as a specific example, we use ultrafast pump-probe and 2DES as a powerful tool for uncovering the electronic properties of quadrupolar molecules in solution and for exploring how electronic couplings and electron delocalization across the molecule largely suppress the vibronic coupling to ubiquitous high-frequency carbon backbone modes. Importantly, the spectroscopic results serve as a crucial benchmark for the quantum-chemical modelling of such vibronic couplings and the resulting coherent quantum dynamics.

The information obtained about squaraine molecules in solution is then used to explore the physical properties of J-aggregated squaraine thin films. Two-quantum 2DES allows us to quantify the blueshift of the two-exciton states in the investigated thin films that lies at the origin of the optical nonlinearity of such aggregated thin films. An analysis of 2DES maps of J-aggregated squaraines on a gold substrate reveals the pronounced effect of near-field couplings between the J-aggregate excitons and surface plasmon polaritons at the gold/thin film interface on the delocalization of the exciton wavefunction. The resulting narrow exciton lineshape makes squaraine-based J-aggregates highly interesting candidates for exploring strong exciton-plasmon couplings. First angle-resolved 2DES studies of a hybrid structure comprising a gold nanoslit grating covered with a 10-nm thick J-aggregate thin film provide evidence for polariton formation and reveal pronounced oscillations of 2DES diagonal and cross peaks during the coherence time of the polariton excitations. A detailed analysis of the 2DES maps and their waiting time dependence shows that the majority of these oscillations are caused by a long-range and coherent transport of excitons across different regions of the sample driven by the plasmon field.

These experiments have been made possible by different technological advances in 2DES: (i) the time resolution of the 2DES experiments has been improved to better than 10 fs using high-repetition rate home-built noncollinear optical parametric amplifiers. This time resolution is shorter than the electronic and vibronic coherence time of the investigated material, even and room temperature. Importantly, it is shorter than the vibrational period of ~ 20 fs of the ubiquitous and functionally relevant carbon backbone vibrations of organic semiconductors allowing for monitoring these vibrations and their couplings to electronic excitations in the time domain; (ii) the spectral bandwidth of the employed NOPA pulses is large enough to fully cover all relevant optical excitations of the material;[22] (iii) advancements in pump laser technology and sensitive high-speed detection offer high signal-to-noise ratios approaching the shot noise limit and reduce data acquisition times and excitation fluences; (iv) the use of an inherently phase-stable inline interferometer based on birefringent wedges largely facilitates the acquisition of absorptive 2DES with (sub-)10 fs time resolution in a partially collinear pump-probe geometry. Recent implementations of phase cycling in such a geometry[36] facilitate the isolation of rephasing and nonrephasing contributions to 2DES, the acquisition of zero- and double-quantum spectra and the selection of specific quantum pathways, of crucial importance for investigating coherent couplings and many-body interactions in quantum materials; (v) simulations of 2DES maps using approaches that are based on response functions and direct time-integration of the Liouville-von-Neumann equation are indispensable tools for gaining deep microscopic insight into the underlying quantum dynamics.

All reported experiments in this chapter have been performed with limited spatial resolution on ensembles of nanostructures. We believe that the recent implementation of phase cycling schemes in TWINS interferometers opens up an exciting path towards 2DES microscopy[66, 67, 70] with high spatial resolution and towards 2DES studies of single nanostructures. These phase-cycling capabilities also make it attractive to further advance combinations of multidimensional and photoelectron emission spectroscopies.[66, 156] The adaptation of recent advances in light-field synthesis[157] may be crucial for further improvements in time resolution in 2DES. Finally, we think bringing together ab-initio quantum simulations[158, 159] and multidimensional coherent spectroscopy may be a particularly fruitful and rewarding path towards improving our current understanding of energy and charge transfer processes in nanostructures.

## 8. Acknowledgments

We wish to thank all past and present members of the Ultrafast Nano-Optics group in Oldenburg for making this work possible. Very special thanks are due to Dr. A. De Sio for her invaluable contributions to the field of multidimensional spectroscopy and beyond. The work reported in this chapter could not have been done without the help of Dr. T. Quenzel, Dr. A. Wöste, Dr. M. Gittinger, D. Lünemann, K. Winte, S. Stephan, R. Angermann and Prof. J.-H. Zhong. We thank M. Schiek, M. F. Schumacher and A. Lützen for providing the squaraine material. We also thank Yu. Zhang, F. Zheng, T. Frauenheim and S. Tretiak for theoretical support.

We acknowledge financial support from Deutsche Forschungsgemeinschaft (SFB1372/2-Sig01 "Magnetoreception and Navigation in Vertebrates", INST 184/163-1, INST 184/164-1, Li 580/16-1, and DE 3578/3-1). We also acknowledge financial support from the Niedersächsische Ministerium für Wissenschaft und Kultur (DyNano and Wissenschaftsraum ElLiKo) and the Volkswagen Foundation (SMART).